\documentclass[12pt,a4paper,titlepage]{report}
\usepackage[centertags]{amsmath}
\usepackage{amsfonts}
\usepackage{amssymb}
\usepackage{amsthm}
\usepackage{bbm}
\usepackage{newlfont}
\usepackage{a4}
\usepackage{enumerate}
\def\nonumchapter#1{%
    \chapter*{#1}
    \addcontentsline{toc}{chapter}{#1}}
\def\prefacesection#1{%
    \chapter*{#1}
    }
\newlength{\defbaselineskip}
\newcommand{\setlinespacing}[1]%
           {\setlength{\baselineskip}{#1 \defbaselineskip}}

\newcommand{\aut}{\mathop{\mathrm {Aut} }\nolimits}
\newcommand{\tr}{\mathop{\mathrm {tr} }\nolimits}
\newcommand{\g}{{\cal G}}
\newcommand{\el}{{\cal L}}
\newcommand{\es}{{\cal S}}

\newcommand{\er}{{\cal R}}

\newcommand{\q}{\quad}
\newcommand{\pe}{{\cal P}}
\newcommand{\I}{{\cal I}}
\newcommand{\E}{{\cal E}}

\newcommand{\qe}{{\cal Q}}

\newcommand{\Ng}{\cal N(\cal G)}

\renewcommand{\epsilon}{\varepsilon}
\newcommand{\ep}{\varepsilon}
\newcommand{\la}{\lambda}

\renewcommand{\rho}{\varrho}
\renewcommand{\phi}{\varphi}

\newcommand{\Com}{{\mathbb C}}

\newcommand{\Z}{\mathbb{Z}}

\newcommand{\set}[1]{\left\{#1\right\}}

\newcommand{\wt}{\widetilde}
\theoremstyle{definition}
\newtheorem{defn}{Definice}[section]
\newtheorem{thm}[defn]{Theorem}
\newtheorem{lemma}[defn]{Lemma}
\newtheorem{tvr}[defn]{Proposition}
\newtheorem{cor}[defn]{Corollary}
\theoremstyle{remark}
\newtheorem{remark}{Remark}
\newtheorem{example}{Example}
\begin{document}

\begin{titlepage}
\begin{center}
\Large{CZECH TECHNICAL UNIVERSITY IN PRAGUE} \\\normalsize Faculty of Nuclear Sciences
and Physical Engineering
\end{center}
\addvspace{150pt}
\begin{center}
\LARGE{\bf{DIPLOMA THESIS}}
\end{center}
\addvspace{50pt}
\begin{center}
\LARGE SOLUTION OF CONTRACTION EQUATIONS FOR THE PAULI GRADING OF $sl(3,\Com )$
\end{center}
\addvspace{50pt}
\begin{center}
\Large Ji\v{r}\'{\i} Hrivn\'ak \bigskip\bigskip\\ \normalsize Supervisor: Prof. Ing. {J.}
Tolar, DrSc.
\end{center}
\addvspace{80pt}
\begin{center}
\large April 22, 2003
\end{center}
\end{titlepage}
\begin{titlepage}
\begin{flushleft}
\addvspace{350pt}
\bigskip
I declare that I wrote this diploma thesis independently using the listed references.
\end{flushleft}
\begin{flushright}
\addvspace{30pt}
\bigskip
Ji\v{r}\'{\i} Hrivn\'ak
\end{flushright}
\end{titlepage}

\setlinespacing{1.1}
\prefacesection{Acknowledgements}
\thispagestyle{empty} \pagenumbering{roman} I would like to thank prof. Miloslav
Havl\'{\i}\v{c}ek and doc. Edita Pelantov\'a for advice and ideas on topics of Lie
gradings and graded contractions, prof. Ji\v{r}\'{\i} Patera for advice, ideas and care
during my study stay at Centre de recherches math\'ematiques, Universit\'e de Montr\'eal,
where this work was partially written.

Especially, I would like to thank prof. Ji\v{r}\'{\i} Tolar for his kind supervision
during the past three years.

\newpage
\addcontentsline{toc}{chapter}{Contents} \pagenumbering{arabic} \setlinespacing{1.05}
\tableofcontents
\newpage
\setlinespacing{1.25} \nonumchapter{Introduction}  Lie algebra $A_2 =sl(3,\Com)$ is the
second lowest dimensional among the classical Lie algebras $A_n$. It, as well as its real
forms, have found numerous applications in physics. Also its subalgebras are of great
interest. There are two simple subalgebras, $o(3)$ and $sl(2,\Com)$; Other subalgebras
gained importance later and all of them were classified. The ubiquity of $sl(3,\Com)$
leads to an interesting question of its relation to other Lie algebras (excluding
homomorphisms). One type of this relation are contractions, which were introduced by
Wigner and In\"on\"u in 1953. Here we are interested in contractions of $sl(3,\Com)$
which lead to Lie algebras of the same dimension 8. It turns out that the outcomes of
various kinds of contractions are numerous, but at present not all of them are known,
even in the case of $sl(3,\Com)$. The most general approach allowed classification of Lie
algebras in dimensions 2, 3, 4, 5 and 6 in \cite{sherp}. Thus, the set of all Lie
algebras of dimension 8 is still unknown. The method of graded contractions allows us to
partially fill this gap.

The goal of describing all graded contractions of $sl(3,\Com)$ has a lot of merit, and
undoubtedly can be reached within a relatively short time. The starting point for
achieving this goal are 4 fine gradings of $sl(3,\Com)$ which are known \cite{HPP5}.
Well-known is the toroidal grading, which decomposes $sl(3,\Com)$ into 6 one-dimensional
subspaces (root spaces) and one two-dimensional subspace (the Cartan subalgebra). All the
graded contractions for this grading were found in \cite{W1} and more recently in
\cite{W2}. The other 3 gradings and the corresponding graded contractions will
undoubtedly yield many other Lie algebras which are non-isomorphic to those in \cite{W1}.
Among three remaining gradings, the grading by generalized Pauli matrices \cite{PZ2} is
considered in this work. It is distinguished from the others: it has very few coarsenings
which are intermediate between the original $sl(3,\Com)$ and the finest Pauli grading.
For that reason the solution of the system of contraction equation is the most difficult
one of the four cases since the method in \cite{W2} is ineffective here. Another
interesting outcome is the general result for gradings of Lie algebras $so(N+1)$ in
\cite{Spanien}. Unfortunately, we found a straightforward generalization impossible, even
for the case $sl(3,\Com)$.

For the necessary explicit evaluation of the solutions the symmetry group of the Pauli
grading \cite{HPP4} has been employed in this work. The method of using the symmetry
group and reducing the case by case analysis is, however, developed generally and then
applied to our concrete case. This method was already foreseen in \cite{jiricek}, and
since the symmetry group is in our case isomorphic to a finite matrix group, we make use
of \cite{xpipa}.

The facts and definitions of gradings and graded contractions are stated in the first and
the second chapter, then the symmetry group and its action on the solutions and equations
is introduced in the third chapter. The evaluation of the solutions is presented in the
fourth chapter. The final results in Appendix will serve as an entry to a further
analysis of desired Lie algebras which are graded contractions of $sl(3,\Com)$
corresponding to the Pauli grading.

\newpage

\chapter{Lie gradings}
\section{Basic definitions} Let us first state the basic definitions of Lie gradings. We
consider a Lie algebra $\el$ over the field of complex numbers $\Com$. We shall focus on
finite-dimensional cases, so let the dimension of $\el$ be finite. A decomposition of
this algebra into a direct sum of its subspaces $\el_i, i\in I$
\begin{equation}\label{gra}
\el= \bigoplus _{i \in I} \el_i
\end{equation}
is called a {\bf grading} of Lie algebra $\el$, when the following property holds
\begin{equation}\label{grad}
  (\forall i, j \in I )(\exists k \in I)( [ \el_{i},
\el_{j} ] \subseteq  \el_{k}),
\end{equation}
where $I$ is an index set, and we denote
\begin{equation}\label{HH23}
  [ \el_{i}, \el_{j} ]:= \left\{[x,y] \big|x \in \el_i,y \in \el_j  \right\}.
\end{equation}
Subspaces $\el_i, i \in I $ are then called {\bf grading subspaces}.

Gradings of Lie algebra $\el$ are closely related to the group of automorphisms $\aut
\el$. Let us recall that a regular linear mapping $g$ acting on $\el$, i.e. $g \in
GL(\el)$, is an {\bf automorphism} of $\el$ if
\begin{equation}\label{automo}
g \,[X,Y]=[gX,gY]
\end{equation}
holds for all $X,Y \in \el$. The relationship between two gradings can be described in
the following way: if $\Gamma : \el=\bigoplus _{i \in I} \el_i$ is a grading of $\el$,
then for an arbitrary automorphism $g \in \aut \el$  $$\tilde{\Gamma} :\el=\bigoplus _{i
\in I} g(\el_i) $$ is also a grading of $\el$. We call such gradings $\Gamma$ and
$\tilde{\Gamma}$ {\bf equivalent}.

\bigskip
Grading $\Gamma : \el=\bigoplus _{i \in I} \el_i$ is a {\bf refinement} of grading
$\tilde{\Gamma} : \el=\bigoplus _{j \in J} \tilde{\el_j}$ if for each $i \in I$ exists $j
\in J$ such that $\el_i \subseteq \tilde{\el_j} $. Refinement is called {\bf proper} if
the cardinality of $I$ is greater than the cardinality of the set $J$.
 Grading which cannot be properly refined is called {\bf fine}.
 If all grading subspaces are one-dimensional then the grading is called
 {\bf finest}.

The property~(\ref{grad}) defines a binary operation on the set $I$. If $[\el_i,\el_j]=\{
0 \}$ holds, we can choose an arbitrary $k$. It is proved in \cite{Pat1}  that this index
set $I$ with this operation can always be embedded into an {\it Abelian group} $G$; we
are going to denote  the operation additively as $+$ and we have
\begin{equation}\label{grupa}
  [\el_i,\el_j]\subseteq \el_{i+j}\,, \q \mbox{where  }\,i,j,i+j \in G .
\end{equation}
We say that the Lie algebra is graded by group $G$ or it is {\bf {\it G}-graded}. Group
$G$ is called a {\bf grading group}.

\bigskip
\section{The group $\aut\el$ and gradings}

This section describes the process of obtaining gradings and the correspondence between
automorphisms and gradings. Let $g \in \aut \el$ be a {\it diagonable} automorphism. Let
$X$ and $Y$ be eigenvectors of $g$ corresponding to (of course non--zero) eigenvalues
$\la$ and $\mu$, e.g. $$ g\, X = \la X, \quad g\,Y= \mu Y.$$ Then we have $$g
\,[X,Y]=[gX,gY]=\la \mu [X,Y],$$ hence $[X,Y]$ is either eigenvector corresponding to the
eigenvalue $\la \mu$ or zero vector. It follows that $$[\el_\la ,\el_\mu] \subseteq
\el_{\la\mu}.$$ This means that the decomposition of $\el$ into the direct sum of
eigenspaces of diagonable automorphism $g$ is the grading of $\el$ :
\begin{equation}\label{}
\el = \bigoplus_{i\in I} \mbox{Ker} (g - \lambda_i \, {\mbox{id}}),
\end{equation}
where $I$ is the set indexing all eigenvalues of $g$. If we take another diagonable
automorphism $h$, which commutes with $g$, then there exist common eigenvectors (and
eigenspaces) which determine the same or a finer grading. In this way every set of
diagonable and mutually commuting automorphisms $g_1,g_2,\dots,g_m\ \in \aut \el$
determines some grading.

The maximal set of diagonable and mutually commuting automorphisms is in fact a subgroup
of $\aut \el$ called {\bf MAD-group} ({\bf m}aximal {\bf A}belian group of {\bf
d}iagonable automorphisms). Conversely, each given grading (\ref{grad}) determines a
subgroup Diag$\,\Gamma \subset \aut {\mathcal L}$ containing all automorphisms $g \in
GL({\mathcal L}),$ which preserve $\Gamma$, $g({\mathcal L}_i) = {\mathcal L}_i$, and are
{\bf diagonal}, $$ g x = \lambda_i x  \q  \mbox{for all}\: x \in {\el}_i, \, i \in I,$$
where $\lambda_i \neq 0$ depends only on $g$ and $i \in I$. In \cite{Pat1} an important
theorem has been proved which for all simple Lie algebras classifies all their possible
fine gradings.

\begin{thm}{\label{thm1}}
Let $\el$ be a finite-dimensional simple Lie algebra over an algebraically closed field
of characteristic zero. Then the grading $\Gamma$ is fine if and only if Diag$\,\Gamma$
is equal to some MAD-group.
\end{thm}
\bigskip

In this way, the problem of classification of all fine gradings of simple Lie algebras is
converted to a classification of all MAD-groups in $\aut \el$. Since classical simple Lie
algebras are subalgebras of $gl(n,\Com)$, we will first investigate MAD-groups in $\aut
gl(n , \Com)$. Adding supplementary conditions one can obtain  MAD-groups of other
classical algebras.

Automorphisms of $gl(n,\Com)$ can be written as a combination of {\it inner} and {\it
outer} automorphisms. For all $X \in gl(n,\Com)$,
\begin{quote}
 {\bf inner automorphisms} have the general form
\begin{equation}\label{vaut}
 \mbox{Ad}_{A} \, X = A^{-1}XA \q
    \mbox{where} \q  A \in GL(n,\Com);
\end{equation}
 {\bf outer automorphisms} have the general form
\begin{equation}\label{vnaut}
    \mbox{Out}_{A} \, X = -
(A^{-1}XA)^{T} =
          \mbox{Out}_{I} \mbox{Ad}_{A} X,\q
   \mbox{where} \q  A \in GL(n,\Com).
\end{equation}

\end{quote}
We further convert the characteristics of automorphisms in MAD-groups to the
characteristics of corresponding matrices in $GL(n,\Com)$. The properties of all inner
and outer automorphisms are summarized in the following lemma  \cite{HPP1}:
\begin{tvr}\label{auto}
Let $A,B,C \in GL(n,\Com)$. Then the following holds
\begin{enumerate}[(1)]
\item Ad$_{A}$ is diagonable automorphism iff a matrix $A$ is diagonable.
\item Inner automorphisms commute, i.e.
 Ad$_{A}$ Ad$_{B}$ =Ad$_{B}$ Ad$_{A}$,
iff there exists $q \in \Com$ such that
\begin{equation} \label{02}
AB = q BA, \quad \text{where} \,\,\, q \,\,\, \mbox{satisfies} \quad
q^{n} = 1.
\end{equation}
\item  Out$_{C}$ is diagonable iff
    matrix $C(C^{T})^{-1}$ is diagonable.
\item  Inner and outer automorphisms commute, i.e.
 Ad$_{A}$ Out$_{C}$ =Out$_{C}$ Ad$_{A}$ iff
$$ACA^{T} = r C\,.$$ Since Ad$_{\alpha A}$  = Ad$_{A}$ for $\alpha \neq 0$, the number
$r$ can be normalized to unity.
\end{enumerate}
\end{tvr}

\section{The Pauli grading of $sl(n,\Com)$} Since this section deals only
with MAD-groups without outer automorphisms, let us assume that $\g \subset \aut
gl(n,\Com)$ is MAD-group without outer automorphism. Let us consider the set of
corresponding matrices in $GL(n,\Com)$
\begin{equation}\label{Ad}
  G:=\set{A \in GL(n,\Com)~|\,\mbox{Ad}_{A}\in \g }
\end{equation}
and we have indeed
\begin{equation}\label{Mad}
  \g= \mbox{Ad}\, G
:=\set{\mbox{Ad}_{A} ~|\,A\in G }.
\end{equation}
 According to (1) and (2) of lemma
\ref{auto} $G$ is a maximal set of diagonable matrices in $GL(n,\Com)$ such that
$AB=q(A,B)BA$ for all $A,B \in G$. This leads us to the following definition: a subgroup
of diagonable matrices $G \subset GL(n,\Com)$ is called {\bf Ad-group} if
\begin{enumerate}[(i)]
\item  For all $A,B \in G$ the commutator
$q(A,B) = ABA^{-1}B^{-1}$ is a non zero multiple of identity matrix, i.e. it belongs to
the center $Z = \{\alpha I_{n} \vert \alpha \in \Com \setminus \{0\}\} \subset
GL(n,\Com)$
\item  $G$ is maximal, $(\forall M
\notin G)(\exists A \in G)( q(A,M) \notin Z)$.
\end{enumerate}
To each MAD-group in $\aut gl(n,\Com)$ without outer automorphisms there corresponds
(according to (\ref{Ad})) an Ad-group in $GL (n,\Com) $ and similarly {\it vice versa},
according to the formula (\ref{Mad}) to each Ad-group in $GL (n,\Com) $ corresponds a
MAD-group without outer automorphisms.

In order to describe Ad-groups in $GL (n,\Com) $ we introduce the following notation.
Subgroup in $GL(n,\Com)$ containing all regular diagonal matrices is denoted $D(n)$. We
define also $k \times k$ matrices

\begin{equation}\label{maq}
Q_{k}=
\mbox{diag}(1,\omega_k,\omega_{k}^2,\ldots, \omega_{k}^{k-1}),
\end{equation}
 where
$\omega_{k} = \exp (2 \pi i/k)$, and matrix
\begin{equation}\label{map}
P_{k}=
\begin{pmatrix}
0&1&0& \cdots &0&0\\
0&0&1& \cdots &0&0\\
\vdots&&&\ddots & &\\
0&0&0&\cdots& 0&1\\
1&0&0&\cdots &0&0\\
\end{pmatrix}.
\end{equation}
For $k=1$ we set $Q_1 = P_1 = (1)$. For matrices $P_k,Q_k$
\begin{equation}\label{pq}
  P_{k}Q_{k}=  \omega_{k} Q_{k}P_{k}\,
\end{equation}
holds, hence they satisfy (\ref{02}) with $n=k,\,q=\omega _k$. The finite subgroup of
$GL(k,\Com)$ of order $k^3$ defined as
\begin{equation}\label{Pauli}
\Pi_{k} = \{ \omega_{k}^{l} Q_{k}^{i} P_{k}^{j} \,\, \vert \,\,
                     i,j, l = 0,1,\ldots,k-1 \}
\end{equation}
is called the {\bf Pauli group}. Ad-groups in $GL(n,\Com)$ are classified by the
following theorem proved in \cite{HPP2}.
\begin{thm}\label{thm2}
  $G \subset GL(n,\mathbb{C})$ is an Ad-group if and only if
$G$ is conjugated to one of the finite groups $$\Pi_{\pi_1} \otimes \cdots
\otimes
     \Pi_{\pi_s} \otimes D(n/\pi_{1} \ldots \pi_{s}), $$
where  $\pi_{1},\ldots,\pi_{s}$ are powers of primes and their product
$\pi_{1} \ldots \pi_{s}$ divides $n$, with the exception of the case
$\Pi_{2} \otimes \cdots \otimes
     \Pi_{2} \otimes D(1).$
\end{thm}
\begin{remark}
Subgroups $H_1,H_2\subset G$ are conjugated if there exists $g \in G $ such that $H_1=g
H_2 g^{-1}=\set{ghg^{-1}~|\,h\in H_2 }$. Since
$\mbox{Ad}_A=\mbox{Ad}_g\mbox{Ad}_B\mbox{Ad}^{-1}_g$ iff $A=gBg^{-1}$ holds for any
$A,B,g \in GL(n,\Com),$ we see that conjugated Ad-groups correspond to conjugated
MAD-groups which give equivalent gradings.
\end{remark}

\bigskip

We are interested in the case when an Ad-group is equal exactly to $\Pi_n$. The
corresponding MAD-group in $\aut gl(n,\Com)$ is clearly of order $n^2$ (matrices which
differ only by a multiplier give equal automorphism due to (\ref{vaut}))
\begin{equation}\label{adpi}
\mbox{Ad}\,\Pi_n = \{ \text{Ad}_{Q^{i}P^{j}} \vert (i,j) \in
   \mathbb{Z}_{n} \times \mathbb{Z}_{n} \}.
\end{equation}
The fine grading of $gl(n,\Com)$ corresponding to this MAD-group is, according to
\cite{PZ2}, given by
\begin{equation} \label{Grgl}
gl(n,\Com) = \bigoplus_{(r,s) \in
      \mathbb{Z}_{n} \times \mathbb{Z}_{n} }{\el}_{rs},
\end{equation}
where $\el_{rs}:=\{X_{rs}\}_{lin}$ and
\begin{equation}\label{xrs}
  X_{rs}={Q_n}^r{P_n}^s\,.
\end{equation}
This grading is in fact finest, i.e. all $n^2 = \mbox{dim}\,gl(n,\Com)$ subspaces are
one-dimensional. We can easily check that (\ref{Grgl}) is indeed a grading by
verification of the property (\ref{grad}); (henceforth the explicit notation of dimension
$P_n,Q_n$ and for algebraic operations $mod\,\, n$ will be omitted)
\begin{equation}\label{komutator}
[X_{rs}, X_{r' s'}] = Q^{r} P^{s} Q^{r'} P^{s'} -
     Q^{r'} P^{s'} Q^{r} P^{s} =(\omega^{sr'}- \omega^{rs'})
      X_{r+r',s + s'}\,
\end{equation}
where relation
\begin{equation}\label{komu}
  P^sQ^r=\omega^{sr}Q^rP^s
\end{equation}
following from (\ref{pq}) was used. Hence we have that our grading group $G$ is equal to
the additive Abelian group $ \mathbb{Z}_{n} \times \mathbb{Z}_{n}$ with addition
componentwise $(mod\,\,n)$. For us it is important to notice  that the result of the
computation (\ref{komutator}) is never the generator $X_{00}$ for $(r,s)\neq (0,0)$ and
$(r',s')\neq (0,0)$; namely $r+r' \,\,(mod\,n) = 0 ,\,\,s+s' \,\,(mod\,n) = 0\,$ lead to
$$[X_{rs}, X_{-r -s}] = 0\cdot X_{00}=\{0\}.$$ Since for all matrices $X_{rs}$ except
$X_{00}$ $$\tr X_{rs}=0,\,\,\,(r,s)\neq (0,0)$$  holds, we can state that these $n^2-1$
matrices yield a grading of $sl(n,\Com)$:
\begin{equation}\label{slg}
  sl(n,\Com) =
\bigoplus_{(r,s)\in \Z_n \times \Z_n \setminus (0,0) }{\el}_{rs}
\end{equation}
This grading of $sl(n,\Com)$ is called the {\bf Pauli grading}.

\section{Four gradings of $sl(3,\Com)$}

According to \cite{HPP1} $\aut sl(3,\Com)$ has four non-conjugate MAD-groups and
therefore four inequivalent fine gradings. They will be listed according to
\cite{HPP5},\cite{HPP6}. First we list the grading group $G$ and then the corresponding
grading $\Gamma$. The symbol for linear span in the notation involving explicit matrices
is omitted.

\noindent
$G_1 = \Z_3 \times \Z_3$
\noindent
\begin{align}
\Gamma_1: sl(3,\Com) &=  \el_{00}\oplus \el_{10}\oplus \el_{01} \oplus \el_{11}\oplus
\el_{-1-1} \oplus \el_{0-1}\oplus \el_{-10}\\
                     &=\begin{pmatrix}
        a&0&0\\
        0&b&0\\
        0&0&-a-b
        \end{pmatrix}\oplus
\begin{pmatrix}
        0&1&0\\
        0&0&0\\
        0&0&0
\end{pmatrix}\oplus
\begin{pmatrix}
        0&0&0\\
        0&0&1\\
        0&0&0
\end{pmatrix}\oplus
\begin{pmatrix}
        0&0&1\\
        0&0&0\\
        0&0&0
        \end{pmatrix}\oplus\nonumber \\ \nonumber
& \oplus \begin{pmatrix}
        0&0&0\\
        0&0&0\\
        1&0&0
        \end{pmatrix}\oplus
\begin{pmatrix}
        0&0&0\\
        0&0&0\\
        0&1&0
        \end{pmatrix}\oplus
\begin{pmatrix}
        0&0&0\\
        1&0&0\\
        0&0&0
        \end{pmatrix} \nonumber
\end{align}
\noindent $G_2 = \Z_2 \times \Z_2 \times \Z_2$
\begin{align}
\Gamma_2: sl(3,\Com) &=  \el_{001}\oplus \el_{111}\oplus \el_{101} \oplus \el_{011}\oplus
\el_{110} \oplus \el_{010}\oplus \el_{100} \\
                     &= \begin{pmatrix}
        a&0&0\\
        0&b&0\\
        0&0&-a-b
        \end{pmatrix}\oplus
\begin{pmatrix}
        0&1&0\\
        1&0&0\\
        0&0&0
        \end{pmatrix}\oplus
\begin{pmatrix}
        0&0&1\\
        0&0&0\\
        1&0&0
        \end{pmatrix}\oplus
\begin{pmatrix}
        0&0&0\\
        0&0&1\\
        0&1&0
        \end{pmatrix}\oplus\nonumber \\
         & \oplus\begin{pmatrix}
        0&-1&0\\
        1&0&0\\
        0&0&0
        \end{pmatrix}\oplus
        \begin{pmatrix}
        0&0&0\\
        0&0&-1\\
        0&1&0
        \end{pmatrix}\oplus
\begin{pmatrix}
        0&0&-1\\
        0&0&0\\
        1&0&0
        \end{pmatrix} \nonumber
\end{align}
\noindent $G_3 = \Z_8$
\begin{align}
 \Gamma_3: sl(3,\Com) &=  \el_{0}\oplus \el_{1}\oplus \el_{2} \oplus \el_{3}\oplus \el_{4} \oplus \el_{5}\oplus \el_{6}\oplus \el_{7} \\
              &=\begin{pmatrix}
        0&0&0\\
        0&1&0\\
        0&0&-1
        \end{pmatrix}\oplus
\begin{pmatrix}
        0&1&0\\
        0&0&0\\
        -1&0&0
        \end{pmatrix}\oplus
\begin{pmatrix}
        0&0&0\\
        0&0&1\\
        0&0&0
        \end{pmatrix}\oplus
\begin{pmatrix}
        0&0&1\\
        1&0&0\\
        0&0&0
        \end{pmatrix}\oplus\nonumber \\
&\oplus\begin{pmatrix}
        2&0&0\\
        0&-1&0\\
        0&0&-1
        \end{pmatrix}\oplus
\begin{pmatrix}
        0&1&0\\
        0&0&0\\
        1&0&0
        \end{pmatrix}\oplus
\begin{pmatrix}
        0&0&0\\
        0&0&0\\
        0&1&0
        \end{pmatrix}\oplus
\begin{pmatrix}
        0&0&1\\
        -1&0&0\\
        0&0&0
        \end{pmatrix}\nonumber
\end{align}
\noindent $G_4 = \Z_3 \times \Z_3 $ \noindent
\begin{align}\label{zkos}
\Gamma_4: sl(3,\Com) &=   \el_{01}\oplus \el_{02}\oplus \el_{10} \oplus \el_{20}\oplus
\el_{11}\oplus \el_{22} \oplus \el_{12}\oplus \el_{21} \\ \label{zkos}
                     &= Q \oplus Q^2 \oplus P \oplus P^2 \oplus PQ \oplus P^2Q^2 \oplus PQ^2 \oplus P^2Q  \nonumber \\
                     &= \begin{pmatrix}
1 & 0 & 0 \\
0 & \omega  & 0 \\
0 & 0 & \omega ^{2}
\end{pmatrix}
\oplus
\begin{pmatrix}
1 & 0 & 0 \\
0 & \omega ^{2} & 0 \\
0 & 0 & \omega
\end{pmatrix}
\oplus
\begin{pmatrix}
0 & 1 & 0 \\
0 & 0 & 1 \\
1 & 0 & 0
\end{pmatrix}
 \oplus
\begin{pmatrix}
0 & 0 & 1 \\
1 & 0 & 0 \\
0 & 1 & 0
\end{pmatrix}
 \oplus \nonumber\\& \oplus
\begin{pmatrix}
0 & \omega  & 0 \\
0 & 0 & \omega ^{2} \\
1 & 0 & 0
\end{pmatrix}
 \oplus
\begin{pmatrix}
0 & 0 & \omega  \\
1 & 0 & 0 \\
0 & \omega ^{2} & 0
\end{pmatrix}
 \oplus
\begin{pmatrix}
0 & \omega ^{2} & 0 \\
0 & 0 & \omega  \\
1 & 0 & 0
\end{pmatrix}
 \oplus
\begin{pmatrix}
0 & 0 & \omega ^{2} \\
1 & 0 & 0 \\
0 & \omega  & 0
\end{pmatrix} \nonumber
\end{align}
where $\omega=e^{\frac{2\pi i}{3}}$ and matrices $P,Q$ are defined by the formulas
(\ref{map}) and (\ref{maq}) for the case $n=3$. The grading $\Gamma_1$ is the Cartan
decomposition of $sl(3,\Com)$ and $\Gamma_4$ is indeed the Pauli grading of $sl(3,\Com)$.

\newpage
\chapter{Graded contractions of Lie algebras}
\section{Definitions and basic properties} Let us state the definition of a graded
contraction. Suppose $\el$ is a Lie algebra graded by the group $G$, i.e. the relations
(\ref{gra}) and (\ref{grupa}) hold. We make use of the following notation determining
which $[\el_i,\el_j]$ is zero or non-zero. This information is encoded in the matrix
$\kappa := (\kappa_{ij})$ defined as
\begin{equation}\label{kappa}
\begin{matrix}
    \kappa_{ij}=0 & \text{if}\q [\el_i,\el_j]=\{0\}, \\
    \kappa_{ij}=1 & \text{if}\q [\el_i,\el_j]\neq \{0\}.
\end{matrix}
\end{equation}
The matrix $\kappa$ is of course symmetric and of order $k \times k$, where $k$ is the
number of grading subspaces.

We define a bilinear mapping $[\,\,,\,\,]_{\gamma}$ on $\el$ (more precisely on the
underlying vector space $V$) by the formula
\begin{equation}\label{kontr}
  [x,y]_{\gamma} := \gamma_{ij} [x,y] \q {\mbox{for all}} \q x \in {\el}_{i}, \ y\in {\el}_{j},\, i,j\in I\,\,\mbox{and}\,\, \gamma_{ij}\in \Com.
\end{equation}
Since we claim the bilinearity of $[\,\,,\,\,]_{\gamma}$, the condition (\ref{kontr})
determines this mapping on the whole $V$. If we introduce the {\bf contraction
parameters} $\ep_{ij}$ via the equation (with no summation implied)
\begin{equation}\label{eps}
  \ep_{ij}:=\gamma_{ij}\kappa_{ij}\,,
\end{equation}
we can as well write for all $x \in {\el}_{i}, \ y\in {\el}_{j},\,i,j\in I$
\begin{equation}\label{kontr2}
  [x,y]_{\gamma} =[x,y]_{\ep}= \ep_{ij} [x,y]
\end{equation}
because if the subspaces $\el_i,\,\el_j$ commute, $[\el_i,\el_j]=\{0\}$, then it is
always $[\el_i,\el_j]_{\gamma}=\{0\}$ independently of $\gamma_{ij}$. Henceforth we will
mostly deal with the variables $\ep$ which have zeros on the {\it irrelevant positions}
such that $[\el_i,\el_j]=\{0\}$. If $\el^\ep:=(V,[\,\,,\,\,]_{\ep})$ is a Lie algebra,
then it is called the {\bf graded contraction} of the Lie algebra $\el$. Note that the
contraction preserves a grading because it is also true that
\begin{equation}\label{gra2}
\el^\ep= \bigoplus _{i \in G} \el_i
\end{equation}
is a grading of $\el^\ep$.

The two conditions (and their solutions) which the parameters $\ep_{ij}$ must fulfill for
$\el^\ep$ to become a Lie algebra will be considered in the following sections. The first
condition of antisymmetry of $[\,\,,\,\,]_{\ep}$ immediately gives
\begin{equation}\label{antisd}
  \ep_{ij}=\ep_{ji}
\end{equation}
Hence each such solution can be written in the form of a {\it symmetric} matrix $\ep
=(\ep_{ij})$ which is called the {\bf contraction matrix}. The validity of the Jacobi
identity is the second condition and it is equivalent to the property: for all
(unordered) triples $i,j,k \in I$
\begin{equation}\label{eqpp} e(i\:j\:k):\, [x,[y,z]_\ep]_\ep + [z,[x,y]_\ep]_\ep +[y,[z,x]_\ep]_\ep
=0\q (\forall x \in \el_i)(\forall y\in \el_j)(\forall z \in \el_k).
\end{equation}
is satisfied. Each $ e(i\:j\:k),\,i,j,k \in I$ is then called a {\bf contraction
equation}. Using the Jacobi identity in $\el$
\begin{equation}\label{eqpp2}  [x,[y,z]] + [z,[x,y]] +[y,[z,x]]
=0\q (\forall x \in \el_i)(\forall y\in \el_j)(\forall z \in \el_k)
\end{equation}
one can rewrite (\ref{eqpp}) in the form
\begin{align*} e(i\:j\:k):\,&(\ep_{i,j+k}\ep_{jk}-\ep_{k,i+j}\ep_{ij})[x,[y,z]] +(\ep_{j,k+i}\ep_{ki}-\ep_{k,i+j}\ep_{ij})[y,[z,x]]
=0 \\ &(\forall x \in \el_i)(\forall y\in \el_j)(\forall z \in \el_k).
\end{align*}
In special cases, when there exist for example $x'\in\el_i,\,y'\in\el_j,\,z'\in\el_k$
such that $[x',[y',z']]$ and $[y',[z',x']]$ are linearly independent then equation
(\ref{eqpp}) is in this case equivalent to 2 two-term equations
\begin{equation}\label{eqpp3} e(i\:j\:k):\, \ep_{i,j+k}\ep_{jk}=\ep_{k,i+j}\ep_{ij}=\ep_{j,k+i}\ep_{ki}.
\end{equation}
\begin{example} Let us write the contraction equation $e((01)(10)(31))$ for the Pauli grading of
$sl(5,\Com)$:
\begin{equation*} e((01)(10)(31)):\, [x,[y,z]_\ep]_\ep + [z,[x,y]_\ep]_\ep +[y,[z,x]_\ep]_\ep
=0\q \forall x \in \el_{01}\,\forall y\in \el_{10}\,\forall z \in \el_{31}.
\end{equation*}
Since the subspaces are one-dimensional, we have
\begin{equation*} e((01)(10)(31)):\, [ X_{01},[ X_{10},X_{31}]_\ep]_\ep + [ X_{31},[ X_{01}, X_{10}]_\ep]_\ep +[ X_{10},[ X_{31}, X_{01}]_\ep]_\ep
=0,
\end{equation*} and using (\ref{komutator}), where $\omega=e^{\frac{2\pi i}{5}}$, we obtain a three-term equation
\begin{align*}
e((01)(10)(31)):\,&\ep_{(01)(10)}\ep_{(11)(31)}(\omega-1)(\omega^3-\omega)+\ep_{(10)(31)}\ep_{(41)(01)}(1-\omega)(1-\omega^4)+\\
&+\ep_{(31)(01)}\ep_{(32)(10)}(1-\omega^3)(\omega^2-1)=0.
\end{align*}
\end{example}

\section{Normalization of contraction matrices}

Let us introduce the so called normalization process for contraction matrices. At first
we introduce {\bf componentwise matrix multiplication $\bullet$}, i.e. for two matrices
$A=(A_{ij}),\,B=(B_{ij})$ we define the matrix $C:=(C_{ij})$ by the formula
\begin{equation}\label{componte}
  C_{ij}:=A_{ij}B_{ij} \q\q \mbox{(no summation)}
\end{equation}
and write $C=A\bullet B$.

For the given grading (\ref{gra}) we define also a matrix $\alpha:=(\alpha_{ij})$, where
\begin{equation}\label{alpe}
  \alpha_{ij}=\frac{a_i a_j}{a_{i+j}}\q \mbox{for}\,\, i,j\in I
\end{equation}
and $a_i \in \Com\setminus \{0\}$ for all $i \in I$. The matrix $\alpha$ is then called a
{\bf normalization matrix}. Normalization is a process based on the following lemma:
\begin{lemma}\label{nor1}
Let $\el^\ep$ be a graded contraction of a graded Lie algebra $\el= \bigoplus _{i \in I}
\el_i$. Then $\el^{\wt{\ep}}$, where $\wt{\ep}:=\alpha\bullet\ep$, is for any
normalization matrix $\alpha$ a graded contraction of $\el$ and the Lie algebras
$\el^{\wt{\ep}}$ and $\el^\ep$ are isomorphic, $\el^{\wt{\ep}}\simeq\el^\ep$.
\end{lemma}
\begin{proof}
We define a diagonal regular linear mapping $h\in GL(\el)$ corresponding to a
normalization matrix $\alpha=\big(\frac{a_i a_j}{a_{i+j}}\big)$ by the formula
\begin{equation}\label{diagz}
h x_i = a_i x_i\q i \in I,\,x_i \in \el_i\,.
\end{equation}
Then for all $x\in\el_i,\,y\in\el_j$ the bilinear mapping
${[x,y]_{\wt{\ep}}=\wt{\ep}_{ij} [x,y]}$ and the Lie bracket ${[x,y]_{\ep}=\ep_{ij}
[x,y]}$ satisfy $$[x,y]_{\wt{\ep}}=\wt{\ep}_{ij} [x,y]=\frac{a_i
a_j}{a_{i+j}}\ep_{ij}[x,y]=h^{-1}[h x,h y ]_{\ep}. $$ Hence $\el^{\wt{\ep}}$ is a Lie
algebra and $h$ is an isomorphism between $\el^{\wt{\ep}}$ and $\el^\ep$.
\end{proof}
Practically, we set the matrix $\wt{\ep}$ and then we try to normalize a given matrix
$\ep$ to $\wt{\ep}$, i.e. find such a normalization matrix $\alpha$ which satisfies the
{\it normalization equations} $\wt{\ep}=\alpha\bullet\ep$. In most cases it is possible
to normalize in this way the matrix $\ep$ to the matrix which consists of only $0$'s and
$1$'s.

\section{Contraction system for the Pauli grading of $sl(3,\Com)$} This section contains
contraction equations for the Pauli grading of $sl(3,\Com)$, i.e. such equations for the
variables $\ep_{ij}$ which must be fulfilled in order that $\el^\ep$ be a Lie algebra.
Generally, the set of all these contraction equations is called the {\bf contraction
system} $\es$, the set of its solutions will be denoted $\er(\es)$; for the Pauli grading
of $sl(n,\Com)$ we denote the contraction system $\es_n$. Let us take the Pauli grading
in the form (\ref{zkos}), i.e.
\begin{equation}\label{sl3}
sl(3,\Com)=\el_{01}\oplus \el_{02}\oplus \el_{10}\oplus \el_{20}\oplus \el_{11}\oplus
\el_{22}\oplus \el_{12}\oplus \el_{21}\,,
\end{equation}
where $\el_{ij}=\{X_{ij} \}_{lin}$ and $X_{01}=\{Q\},\, X_{02}=\{Q^2\},\, X_{10}=\{P\},\,
X_{20}=\{P^2\},\, X_{11}=\{PQ\},\, X_{22}=\{P^2Q^2\},\, X_{12}=\{PQ^2\},\,
X_{21}=\{P^2Q\}.$ The grading group is $\Z_3 \times \Z_3$; no subspace is labeled by
$(0,0).$ Let us state the explicit form of matrix $\kappa$ defined by (\ref{kappa}); it
is $8\times 8$ symmetric matrix and we will order the indices as in formula (\ref{sl3}),
i.e. positions $11,\,12,\,13,\,\dots$ are denoted
$(01)(01),\,(01)(02),\,(01)(10),\,\dots$ and on each of these positions zero or unity is
found depending on
$[\el_{01},\el_{01}],\,[\el_{01},\el_{02}],\,[\el_{01},\el_{10}],\,\dots$ We have
\begin{equation}\label{kap}
 \kappa=
 \begin{pmatrix}
   0 & 0 & 1 & 1 & 1 & 1 & 1 & 1 \\
   0 & 0 & 1 & 1 & 1 & 1 & 1 & 1 \\
   1 & 1 & 0 & 0 & 1 & 1 & 1 & 1 \\
   1 & 1 & 0 & 0 & 1 & 1 & 1 & 1 \\
   1 & 1 & 1 & 1 & 0 & 0 & 1 & 1 \\
   1 & 1 & 1 & 1 & 0 & 0 & 1 & 1 \\
   1 & 1 & 1 & 1 & 1 & 1 & 0 & 0 \\
   1 & 1 & 1 & 1 & 1 & 1 & 0 & 0
 \end{pmatrix}.
\end{equation}
We can see immediately that our task of finding all graded contractions for the Pauli
grading of $sl(3,\Com)$ has $24$ relevant contraction parameters. For $\ep\in \er(\es)$
the {\bf number of zeros} among these 24 parameters is denoted by the symbol $\nu(\ep)$.
The symmetric contraction matrix $\ep$ with $24$ variables is of the following general
form
\begin{equation}\label{epsi}
\ep=
 \begin{pmatrix}
   0 & 0 & \ep_{(01)(10)} & \ep_{(01)(20)} & \ep_{(01)(11)} & \ep_{(01)(22)} & \ep_{(01)(12)} & \ep_{(01)(21)} \\
   0 & 0 & \ep_{(02)(10)} & \ep_{(02)(20)} & \ep_{(02)(11)} & \ep_{(02)(22)} & \ep_{(02)(12)} & \ep_{(02)(21)} \\
   \ep_{(01)(10)} & \ep_{(02)(10)} & 0 & 0 & \ep_{(10)(11)} & \ep_{(10)(22)} & \ep_{(10)(12)} & \ep_{(10)(21)} \\
   \ep_{(01)(20)} & \ep_{(02)(20)} & 0 & 0 & \ep_{(20)(11)} & \ep_{(20)(22)} & \ep_{(20)(12)} & \ep_{(20)(21)} \\
   \ep_{(01)(11)} & \ep_{(02)(11)} & \ep_{(10)(11)} & \ep_{(20)(11)} & 0 & 0 & \ep_{(11)(12)} & \ep_{(11)(21)} \\
   \ep_{(01)(22)} & \ep_{(02)(22)} & \ep_{(10)(22)} & \ep_{(20)(22)} & 0 & 0 & \ep_{(22)(12)} & \ep_{(22)(21)} \\
   \ep_{(01)(12)} & \ep_{(02)(12)} & \ep_{(10)(12)} & \ep_{(20)(12)} & \ep_{(11)(12)} & \ep_{(22)(12)} & 0 & 0 \\
   \ep_{(01)(21)} & \ep_{(02)(21)} & \ep_{(10)(21)} & \ep_{(20)(21)} & \ep_{(11)(21)} & \ep_{(22)(21)} & 0 & 0
 \end{pmatrix}.
\end{equation}

The antisymmetry of the new product $[X_{kl},X_{mn}]_N=\ep_{(kl)(mn)}[X_{kl},X_{mn}]$ for
$sl(3,\Com)$ is granted due to the symmetry of the matrix $\ep$. It is sufficient to
verify the Jacobi identities on vectors of a basis; the vectors $X_{ij}, (i,j)\in \Z_3
\times \Z_3\setminus (0,0)$ form the basis of $sl(3,\Com)$ and the Jacobi identities are
in the form
\begin{equation}\label{jacob2}
[X_{ij},[X_{kl},X_{mn}]_N]_N+[X_{mn},[X_{ij},X_{kl}]_N]_N+[X_{kl},[X_{mn},X_{ij}]_N]_N=0.
\end{equation}
These equations should hold for all possible triples of indices $(i j),(k l),(m n)$. It
is clear that for a triple where two indices are identical the equation is automatically
fulfilled; the equation also does not depend on the ordering of the triples. The number
of equations is then equal to a the combination number ${(^{8}_{3})} = 56$. Modifying
(\ref{jacob2}) we obtain
\begin{equation}\label{jacob3}
\ep_{(ij)(k+m,l+n)}\ep_{(kl)(mn)}[X_{ij},[X_{kl},X_{mn}]]+ \q \mbox{cyclically}\q=0,
\end{equation}
where the word cyclically means that the two remaining terms are obtained from the first
one by the substitutions: $(i j)\mapsto (m n),\,(k l) \mapsto (i j),\,(m n)\mapsto (k
l),\,$ and $(i j)\mapsto (k l),\, (k l)\mapsto (m n),\,(m n)\mapsto (i j)$ respectively.
The commutation relations (\ref{komutator}) give the double commutator
\begin{equation}\label{komm}
[X_{ij},[X_{kl},X_{mn}]]=(\omega^{lm}-\omega^{kn})(\omega^{j(k+m)}-\omega^{i(l+n)})X_{i+k+m,j+l+n}\,.
\end{equation}
If we substitute this into (\ref{jacob3}), we get
\begin{equation}\label{kommm}
\bigl[\,
\ep_{(ij)(k+m,l+n)}\ep_{(kl)(mn)}(\omega^{lm}-\omega^{kn})(\omega^{j(k+m)}-\omega^{i(l+n)})+
\q \mbox{cyclically}\q \bigl] X_{i+k+m,j+l+n}=0.
\end{equation}
From (\ref{kommm}) we see that equations for which $i+k+m=0\, \wedge\, j+l+n=0$ holds
will also be automatically fulfilled. This situation arises in eight cases. Hence the
contraction system consists of $48$ equations.

Let us state an example of a computation of (\ref{kommm}) for the given triple, e.g.
$(01)(02)(10)$. The result is
\begin{equation}\label{pr}
[\ep_{(02)(10)}\ep_{(01)(12)}(\omega^2-1)(\omega
-1)+0+\ep_{(10)(01)}\ep_{(02)(11)}(1-\omega )(\omega^2-1)]X_{10}=0.
\end{equation}
The final form of the other equations is similar to this one. All equations for
$sl(3,\Com)$ are in form $MN=PQ$. The whole contraction system $\es_3$ is presented in
Table 1. When writing contraction equations we took into account that $\ep$ is a
symmetric matrix; for example, instead of $\ep_{(10)(01)}$ we equivalently write
$\ep_{(01)(10)}$. Note that for each equation the corresponding triple $(i j)(k l)(m n)$
of subspaces is given. The significance of matrices from $SL(2,\Z_3)$ will be explained
in the next chapter in connection with the symmetry group of the Pauli grading.

$$
\begin{tabular}[b]{cl|cc}\noalign{\parbox[l][21pt][c]{3pt}{}{\it Table 1: The system $\es_3$ of contraction equations for the Pauli grading of
$sl(3,\Com)$ (1/2)}}\hline
  \parbox[l][21pt][c]{3pt}{}Equation number &$      \q\q\q\q\q\q\q\q\q\q\q\q\q\q\q\q\q\q      $&Subspaces&$ SL(2,\Z_3)$ \\ \hline
  \parbox[l][23pt][c]{5pt}{}1 & $\ep_{(02)(10)}\ep_{(01)(12)}-\ep_{(01)(10)}\ep_{(02)(11)}=0$ & \it{(01)(02)(10)} &$\left( \begin{smallmatrix}  1 & 0 \\ 0 & 1\end{smallmatrix}\right)$\\
  \parbox[l][23pt][c]{5pt}{}2 & $\ep_{(02)(20)}\ep_{(01)(22)}-\ep_{(01)(20)}\ep_{(02)(21)}=0$ & \it{(01)(02)(20)} &$\left( \begin{smallmatrix}  2 & 0 \\ 0 & 2\end{smallmatrix}\right)$\\
  \parbox[l][23pt][c]{5pt}{}3 & $\ep_{(02)(11)}\ep_{(01)(10)}-\ep_{(01)(11)}\ep_{(02)(12)}=0$ & \it{(01)(02)(11)} &$\left( \begin{smallmatrix}  1 & 1 \\ 0 & 1\end{smallmatrix}\right)$\\
  \parbox[l][23pt][c]{5pt}{}4 & $\ep_{(02)(22)}\ep_{(01)(21)}-\ep_{(01)(22)}\ep_{(02)(20)}=0$ & \it{(01)(02)(22)} &$\left( \begin{smallmatrix}  2 & 2 \\ 0 & 2\end{smallmatrix}\right)$\\
  \parbox[c][23pt][c]{5pt}{}5 & $\ep_{(02)(12)}\ep_{(01)(11)}-\ep_{(01)(12)}\ep_{(02)(10)}=0$ & \it{(01)(02)(12)} &$\left( \begin{smallmatrix}  1 & 2 \\ 0 & 1\end{smallmatrix}\right)$\\
  \parbox[c][23pt][c]{5pt}{}6 & $\ep_{(02)(21)}\ep_{(01)(20)}-\ep_{(01)(21)}\ep_{(02)(22)}=0$ & \it{(01)(02)(21)} &$\left( \begin{smallmatrix}  2 & 1 \\ 0 & 2\end{smallmatrix}\right)$\\
  \parbox[c][23pt][c]{5pt}{}7 & $\ep_{(10)(21)}\ep_{(01)(12)}-\ep_{(10)(12)}\ep_{(22)(21)}=0$ & \it{(10)(12)(21)} &$\left( \begin{smallmatrix}  1 & 0 \\ 2 & 1\end{smallmatrix}\right)$\\
  \parbox[c][23pt][c]{5pt}{}8 & $\ep_{(20)(21)}\ep_{(11)(12)}-\ep_{(20)(12)}\ep_{(02)(21)}=0$ & \it{(12)(20)(21)} &$\left( \begin{smallmatrix}  2 & 0 \\ 1 & 2\end{smallmatrix}\right)$\\
  \parbox[c][23pt][c]{5pt}{}9 & $\ep_{(10)(22)}\ep_{(02)(11)}-\ep_{(10)(11)}\ep_{(22)(21)}=0$ & \it{(10)(11)(22)} &$\left( \begin{smallmatrix}  1 & 0 \\ 1 & 1\end{smallmatrix}\right)$\\
  \parbox[l][23pt][c]{5pt}{}10 & $\ep_{(20)(22)}\ep_{(11)(12)}-\ep_{(20)(11)}\ep_{(01)(22)}=0$ & \it{(11)(20)(22)} &$\left( \begin{smallmatrix}  2 & 0 \\ 2 & 2\end{smallmatrix}\right)$\\
  \parbox[l][23pt][c]{5pt}{}11 & $\ep_{(01)(20)}\ep_{(10)(21)}-\ep_{(01)(10)}\ep_{(20)(11)}=0$ & \it{(01)(10)(20)} &$\left( \begin{smallmatrix}  0 & 1 \\ 2 & 0\end{smallmatrix}\right)$\\
  \parbox[c][23pt][c]{5pt}{}12 & $\ep_{(02)(20)}\ep_{(10)(22)}-\ep_{(02)(10)}\ep_{(20)(12)}=0$ & \it{(02)(10)(20)} &$\left( \begin{smallmatrix}  0 & 2 \\ 1 & 0\end{smallmatrix}\right)$\\
  \parbox[c][23pt][c]{5pt}{}13 & $\ep_{(01)(21)}\ep_{(22)(12)}-\ep_{(01)(12)}\ep_{(10)(21)}=0$ & \it{(01)(12)(21)} &$\left( \begin{smallmatrix}  0 & 1 \\ 2 & 1\end{smallmatrix}\right)$\\
  \parbox[c][23pt][c]{5pt}{}14 & $\ep_{(02)(21)}\ep_{(20)(12)}-\ep_{(11)(21)}\ep_{(02)(12)}=0$ & \it{(02)(12)(21)} &$\left( \begin{smallmatrix}  0 & 2 \\ 1 & 2\end{smallmatrix}\right)$\\
  \parbox[c][23pt][c]{5pt}{}15 & $\ep_{(01)(22)}\ep_{(20)(11)}-\ep_{(01)(11)}\ep_{(22)(12)}=0$ & \it{(01)(11)(22)} &$\left( \begin{smallmatrix}  0 & 1 \\ 2 & 2\end{smallmatrix}\right)$\\
  \parbox[c][23pt][c]{5pt}{}16 & $\ep_{(02)(22)}\ep_{(11)(21)}-\ep_{(02)(11)}\ep_{(10)(22)}=0$ & \it{(02)(11)(22)} &$\left( \begin{smallmatrix}  0 & 2 \\ 1 & 1\end{smallmatrix}\right)$\\
  \parbox[c][23pt][c]{5pt}{}17 & $\ep_{(20)(11)}\ep_{(01)(10)}-\ep_{(10)(11)}\ep_{(20)(21)}=0$ & \it{(10)(11)(20)} &$\left( \begin{smallmatrix}  1 & 1 \\ 2 & 0\end{smallmatrix}\right)$\\
  \parbox[c][23pt][c]{5pt}{}18 & $\ep_{(20)(22)}\ep_{(10)(12)}-\ep_{(10)(22)}\ep_{(02)(20)}=0$ & \it{(10)(20)(22)} &$\left( \begin{smallmatrix}  2 & 2 \\ 1 & 0\end{smallmatrix}\right)$\\
  \parbox[c][23pt][c]{5pt}{}19 & $\ep_{(20)(21)}\ep_{(10)(11)}-\ep_{(10)(21)}\ep_{(01)(20)}=0$ & \it{(10)(20)(21)} &$\left( \begin{smallmatrix}  2 & 1 \\ 2 & 0\end{smallmatrix}\right)$\\
  \parbox[c][23pt][c]{5pt}{}20 & $\ep_{(20)(12)}\ep_{(02)(10)}-\ep_{(10)(12)}\ep_{(20)(22)}=0$ & \it{(10)(12)(20)} &$\left( \begin{smallmatrix}  1 & 2 \\ 1 & 0\end{smallmatrix}\right)$\\
  \parbox[c][23pt][c]{5pt}{}21 & $\ep_{(11)(21)}\ep_{(02)(12)}-\ep_{(11)(12)}\ep_{(20)(21)}=0$ & \it{(11)(12)(21)} &$\left( \begin{smallmatrix}  1 & 1 \\ 1 & 2\end{smallmatrix}\right)$\\
  \parbox[c][23pt][c]{5pt}{}22 & $\ep_{(22)(21)}\ep_{(10)(12)}-\ep_{(22)(12)}\ep_{(01)(21)}=0$ & \it{(12)(21)(22)} &$\left( \begin{smallmatrix}  2 & 2 \\ 2 & 1\end{smallmatrix}\right)$\\
  \parbox[c][23pt][c]{5pt}{}23 & $\ep_{(22)(12)}\ep_{(01)(11)}-\ep_{(11)(12)}\ep_{(20)(22)}=0$ & \it{(11)(12)(22)} &$\left( \begin{smallmatrix}  1 & 2 \\ 2 & 2\end{smallmatrix}\right)$\\
  \parbox[c][23pt][c]{5pt}{}24 & $\ep_{(22)(21)}\ep_{(10)(11)}-\ep_{(11)(21)}\ep_{(02)(22)}=0$ & \it{(11)(21)(22)} &$\left( \begin{smallmatrix}  2 & 1 \\ 1 &
  1\end{smallmatrix}\right)$\\ \hline
\end{tabular}
$$

$$
\begin{tabular}[b]{cl|cc}\noalign{\parbox[l][21pt][c]{3pt}{}{\it Table 1: The system $\es_3$ of contraction equations for the Pauli grading of
$sl(3,\Com)$ (2/2)}} \hline
  \parbox[l][21pt][c]{3pt}{}Equation number &$      \q\q\q\q\q\q\q\q\q\q\q\q\q\q\q\q\q\q      $&Subspaces& $ SL(2,\Z_3)$ \\ \hline
  \parbox[l][23pt][c]{5pt}{}25 & $\ep_{(10)(11)}\ep_{(01)(21)}-\ep_{(01)(11)}\ep_{(10)(12)}=0$ & \it{(01)(10)(11)} &$\left( \begin{smallmatrix}  1 & 0 \\ 0 & 1\end{smallmatrix}\right)$\\
  \parbox[l][23pt][c]{5pt}{}26 & $\ep_{(20)(22)}\ep_{(02)(12)}-\ep_{(02)(22)}\ep_{(20)(21)}=0$ & \it{(02)(20)(22)} &$\left( \begin{smallmatrix}  2 & 0 \\ 0 & 2\end{smallmatrix}\right)$\\
  \parbox[l][23pt][c]{5pt}{}27 & $\ep_{(11)(12)}\ep_{(01)(20)}-\ep_{(01)(12)}\ep_{(10)(11)}=0$ & \it{(01)(11)(12)} &$\left( \begin{smallmatrix}  1 & 1 \\ 0 & 1\end{smallmatrix}\right)$\\
  \parbox[l][23pt][c]{5pt}{}28 & $\ep_{(22)(21)}\ep_{(02)(10)}-\ep_{(02)(21)}\ep_{(20)(22)}=0$ & \it{(02)(21)(22)} &$\left( \begin{smallmatrix}  2 & 2 \\ 0 & 2\end{smallmatrix}\right)$\\
  \parbox[l][23pt][c]{5pt}{}29 & $\ep_{(10)(12)}\ep_{(01)(22)}-\ep_{(01)(10)}\ep_{(11)(12)}=0$ & \it{(01)(10)(12)} &$\left( \begin{smallmatrix}  1 & 2 \\ 0 & 1\end{smallmatrix}\right)$\\
  \parbox[l][23pt][c]{5pt}{}30 & $\ep_{(20)(21)}\ep_{(02)(11)}-\ep_{(02)(20)}\ep_{(22)(21)}=0$ & \it{(02)(20)(21)} &$\left( \begin{smallmatrix}  2 & 1 \\ 0 & 2\end{smallmatrix}\right)$\\
  \parbox[l][23pt][c]{5pt}{}31 & $\ep_{(01)(21)}\ep_{(10)(22)}-\ep_{(01)(10)}\ep_{(11)(21)}=0$ & \it{(01)(10)(21)} &$\left( \begin{smallmatrix}  1 & 0 \\ 2 & 1\end{smallmatrix}\right)$\\
  \parbox[l][23pt][c]{5pt}{}32 & $\ep_{(02)(20)}\ep_{(22)(12)}-\ep_{(02)(12)}\ep_{(20)(11)}=0$ & \it{(02)(12)(20)} &$\left( \begin{smallmatrix}  2 & 0 \\ 1 & 2\end{smallmatrix}\right)$\\
  \parbox[l][23pt][c]{5pt}{}33 & $\ep_{(11)(21)}\ep_{(02)(10)}-\ep_{(10)(21)}\ep_{(01)(11)}=0$ & \it{(10)(11)(21)} &$\left( \begin{smallmatrix}  1 & 0 \\ 1 & 1\end{smallmatrix}\right)$\\
  \parbox[l][23pt][c]{5pt}{}34 & $\ep_{(22)(12)}\ep_{(01)(20)}-\ep_{(20)(12)}\ep_{(02)(22)}=0$ & \it{(12)(20)(22)} &$\left( \begin{smallmatrix}  2 & 0 \\ 2 & 2\end{smallmatrix}\right)$\\
  \parbox[l][23pt][c]{5pt}{}35 & $\ep_{(20)(21)}\ep_{(01)(11)}-\ep_{(01)(21)}\ep_{(20)(22)}=0$ & \it{(01)(20)(21)} &$\left( \begin{smallmatrix}  0 & 1 \\ 2 & 0\end{smallmatrix}\right)$\\
  \parbox[l][23pt][c]{5pt}{}36 & $\ep_{(10)(12)}\ep_{(02)(22)}-\ep_{(02)(12)}\ep_{(10)(11)}=0$ & \it{(02)(10)(12)} &$\left( \begin{smallmatrix}  0 & 2 \\ 1 & 0\end{smallmatrix}\right)$\\
  \parbox[l][23pt][c]{5pt}{}37 & $\ep_{(22)(21)}\ep_{(01)(10)}-\ep_{(01)(22)}\ep_{(20)(21)}=0$ & \it{(01)(21)(22)} &$\left( \begin{smallmatrix}  0 & 1 \\ 2 & 1\end{smallmatrix}\right)$\\
  \parbox[l][23pt][c]{5pt}{}38 & $\ep_{(11)(12)}\ep_{(02)(20)}-\ep_{(02)(11)}\ep_{(10)(12)}=0$ & \it{(02)(11)(12)} &$\left( \begin{smallmatrix}  0 & 2 \\ 1 & 2\end{smallmatrix}\right)$\\
  \parbox[l][23pt][c]{5pt}{}39 & $\ep_{(20)(22)}\ep_{(01)(12)}-\ep_{(01)(20)}\ep_{(22)(21)}=0$ & \it{(01)(20)(22)} &$\left( \begin{smallmatrix}  0 & 1 \\ 2 & 2\end{smallmatrix}\right)$\\
  \parbox[l][23pt][c]{5pt}{}40 & $\ep_{(10)(11)}\ep_{(02)(21)}-\ep_{(02)(10)}\ep_{(11)(12)}=0$ & \it{(02)(10)(11)} &$\left( \begin{smallmatrix}  0 & 2 \\ 1 & 1\end{smallmatrix}\right)$\\
  \parbox[l][23pt][c]{5pt}{}41 & $\ep_{(01)(20)}\ep_{(11)(21)}-\ep_{(01)(11)}\ep_{(20)(12)}=0$ & \it{(01)(11)(20)} &$\left( \begin{smallmatrix}  1 & 1 \\ 2 & 0\end{smallmatrix}\right)$\\
  \parbox[l][23pt][c]{5pt}{}42 & $\ep_{(02)(22)}\ep_{(10)(21)}-\ep_{(02)(10)}\ep_{(22)(12)}=0$ & \it{(02)(10)(22)} &$\left( \begin{smallmatrix}  2 & 2 \\ 1 & 0\end{smallmatrix}\right)$\\
  \parbox[l][23pt][c]{5pt}{}43 & $\ep_{(11)(21)}\ep_{(02)(20)}-\ep_{(20)(11)}\ep_{(01)(21)}=0$ & \it{(11)(20)(21)} &$\left( \begin{smallmatrix}  2 & 1 \\ 2 & 0\end{smallmatrix}\right)$\\
  \parbox[l][23pt][c]{5pt}{}44 & $\ep_{(22)(12)}\ep_{(01)(10)}-\ep_{(10)(22)}\ep_{(02)(12)}=0$ & \it{(10)(12)(22)} &$\left( \begin{smallmatrix}  1 & 2 \\ 1 & 0\end{smallmatrix}\right)$\\
  \parbox[l][23pt][c]{5pt}{}45 & $\ep_{(20)(12)}\ep_{(02)(11)}-\ep_{(20)(11)}\ep_{(01)(12)}=0$ & \it{(11)(12)(20)} &$\left( \begin{smallmatrix}  1 & 1 \\ 1 & 2\end{smallmatrix}\right)$\\
  \parbox[l][23pt][c]{5pt}{}46 & $\ep_{(10)(22)}\ep_{(02)(21)}-\ep_{(10)(21)}\ep_{(01)(22)}=0$ & \it{(10)(21)(22)} &$\left( \begin{smallmatrix}  2 & 2 \\ 2 & 1\end{smallmatrix}\right)$\\
  \parbox[l][23pt][c]{5pt}{}47 & $\ep_{(01)(22)}\ep_{(20)(12)}-\ep_{(01)(12)}\ep_{(10)(22)}=0$ & \it{(01)(12)(22)} &$\left( \begin{smallmatrix}  1 & 2 \\ 2 & 2\end{smallmatrix}\right)$\\
  \parbox[l][23pt][c]{5pt}{}48 & $\ep_{(02)(21)}\ep_{(20)(11)}-\ep_{(02)(11)}\ep_{(10)(21)}=0$ & \it{(02)(11)(21)} &$\left( \begin{smallmatrix}  2 & 1 \\ 1 & 1\end{smallmatrix}\right)$\\    \hline
\end{tabular}
$$

\chapter{Symmetries and graded contractions}
\section{Symmetry group of a grading}
\subsection{Definitions and statements}
The contraction system for the Pauli grading of $sl(3,\Com)$ is the system of 48
quadratic polynomials in 24 variables. Employing the symmetries, the solution of such
system could be less complicated. The system $\es$ contains transformed Jacobi
identities; therefore we begin with the following consideration. We define the {\bf
symmetry group} $\aut \Gamma$ of a grading
\begin{equation}\label{graff}
  \Gamma : \el=\bigoplus _{i \in I} \el_i
\end{equation}
as such a subgroup of $\aut \el$ which contains automorphisms $g$ with the property
\begin{equation}\label{per1}
  g \el_i=\el_{\pi_g(i)},
\end{equation}
where $\pi_g :I \rightarrow I$ is a permutation of the index set $I$. Thus, a permutation
representation $\Delta_\Gamma$ of the group $\aut \Gamma$ is given on the set $I$,
defined as
\begin{equation}\label{per2}
  \Delta_\Gamma (g):= \pi_g\,.
\end{equation}
The kernel of this representation is the {\bf stabilizer} of $\Gamma$ in $\aut \Gamma$,
\begin{equation}\label{stab}
 \text{Stab}\ \Gamma = \ker \Delta_\Gamma =
\{ g \in \text{Aut}\ {\mathcal L}\ \vert \ g {\mathcal L}_{i}=
  {\mathcal L}_{i}\, \; \forall i \in I \}.
\end{equation}
Hence the stabilizer is a normal subgroup of $\aut \Gamma$ and, according to the
isomorphism theorem for groups, we have
\begin{equation}\label{stab2}
  \text{Aut}\ \Gamma / \text{Stab}\ \Gamma \simeq
  \Delta_\Gamma \ \text{Aut}\ \Gamma.
\end{equation}
This permutation group $\Delta_\Gamma \ \text{Aut}\ \Gamma$ is crucial for solving the
system $\es$. To determine it we now use relation (\ref{stab2}). For fine gradings
corresponding to the MAD-group $\g$ we have $\text{Stab} \,\Gamma = \g$. We define the
{\bf normalizer} of a MAD-group $\g$ as a subgroup
\begin{equation}\label{norm}
\Ng:=\set{ h \in \aut \el ~|\,h^{-1}\g h \subset \g  }.
\end{equation}
Let us take $h \in \Ng$ and the subspace $\el_i$ of the {\it fine} grading (\ref{graff})
corresponding to the MAD-group $\g$. Then for each $f \in \g$ there exists $g \in \g$
such that $h^{-1}fh=g$. Since $g \el_i=\el_i$ holds we see that $f(h\el_i)=h\el_i$. This
means that $h \el_i$ is the eigenspace of $f \in \g$ and $h\el_i=\el_j$ must hold for
some $j\in I$. We have shown the inclusion $\Ng\subset\aut\Gamma$. {\it Vice versa}, $h
\el_i = \el_{\pi_h(i)}$ holds for $h \in \aut\Gamma$. Choosing arbitrary $f \in \g$ we
see that $$f h \el_i = f \el_{\pi_h(i)}=\el_{\pi_h(i)}=h\el_i $$ implies
$h^{-1}fh\el_i=\el_i$ and $h^{-1}fh\in\g$, i.e. $h \in \Ng$. Finally we have
\begin{equation}\label{eqa}
  \Ng=\aut\Gamma
\end{equation}
and we conclude due to (\ref{stab2}):
\begin{cor}\label{corr}
\begin{equation}\label{conc}
\Ng / \g \simeq \Delta_\Gamma\aut\Gamma
\end{equation}
\end{cor}

\subsection{Action of a symmetry group}
We denote the set of {\bf relevant unordered pairs of grading indices} as $\I$, i.e.
\begin{equation}\label{ungrad}
\I := \left\{ i\:j\,\big|\,i,j\in I,\,[\el_i,\el_j]\neq \{0\} \right\},
\end{equation}
where $i\,\,j$ denotes an unordered pair. For the Pauli grading of $sl(n,\Com)$ we denote
this set as $\I_n$. Analyzing relations (\ref{eps}), (\ref{kappa}) and (\ref{komutator}),
we write explicitly
\begin{equation}\label{IN}
\I_n=\left\{\,(ij)(kl)\,\big|\,\,jk-il\neq 0\,(\mbox{mod
}n),\,\,(ij),(kl)\in\Z_n\times\Z_n\setminus\{(0,0)\}\right\}.
\end{equation}
The set of {\bf relevant contraction parameters} $\ep_{ij}$, due to (\ref{antisd}), can
be written as $\E:=\{ \ep_k,\,k \in \I\}$. For a permutation $\pi \in
\Delta_\Gamma\aut\Gamma$ and a contraction matrix $\ep=(\ep_{ij})$, the {\bf action of
$\pi$ on a contraction matrix} $\ep \mapsto \ep^\pi$ is defined by
\begin{equation}\label{aknn}
 (\ep^\pi)_{ij}:=\ep_{\pi (i)\pi(j)}.
\end{equation}
We observe that {\bf the action on variables} $\ep_{ij}\mapsto\ep_{\pi(i)\pi(j)}$ is
really the action on the set of relevant variables $\E$: if
$\ep_{ij}\in\E,\,[\el_i,\el_j]\neq\{0\}$ and $g\in \aut\Gamma,\,\Delta_\Gamma(g)=\pi$,
then $\{0\}\neq g[\el_i,\el_j]=[\el_{\pi(i)},\el_{\pi(j)}]$ and
$\ep_{\pi(i)\pi(j)}\in\E$. Hence the matrix $\ep^\pi$ has zeros on the same irrelevant
positions as the matrix $\ep$.
\begin{remark}\label{remor} Generally for a group $G$ and a set $X\neq\emptyset$ the {\bf left action} $\phi :G\times X\mapsto X$
 and the {\bf right action} $\psi :G\times X\mapsto X$ of the group $G$ on the set $X$
 satisfy
\begin{enumerate}
    \item $\phi(gh,x)=\phi(g,\phi(h,x))\q \phi(e,x)=x \q \mbox{for all}\,\, x \in X,\, g,h\in G $
    \item $\psi(gh,x)=\psi(h,\psi(g,x))\q \psi(e,x)=x \q \mbox{for all}\,\, x \in X,\, g,h\in G
    $,
\end{enumerate}
where $e\in G$ is the unit element. For the action $\theta$ of the group $G$ on the set
$X$ the relation $a \equiv b \Leftrightarrow (\exists g \in G)(\theta(g,a)=b)$ is an
equivalence on $X$ and the equivalence class corresponding to element $a \in X$
\begin{equation}\label{orbe}
\{b\in X \,|\,(\exists g \in G)(b=\theta(g,a))\}
\end{equation}
is called an {\bf orbit} of $a\in X$. We will verify that (\ref{aknn}) is a well-defined
action of the group $\Delta_\Gamma\aut\Gamma$ on the set $\er(\es)$ of the contraction
system solutions. However, we are going to make use of a different definition of
equivalence on the solutions further on.
\end{remark}
In order to verify now that we have defined a correct action on the set of contraction
matrices, we state:
\begin{lemma}\label{nor2}
Let $\el^\ep$ be a graded contraction of a graded Lie algebra $\Gamma : \el=\bigoplus _{i
\in I} \el_i$. Then $\el^{\ep^\pi}$ is for any permutation $\pi \in \Delta_\Gamma \
\text{Aut}\ \Gamma$ a graded contraction of $\el$ and two Lie algebras $\el^{\ep^\pi}$
and $\el^\ep$ are isomorphic, $\el^{\ep^\pi}\simeq \el^\ep$.
\end{lemma}
\begin{proof}
For given $\pi \in \Delta_\Gamma \ \text{Aut}\ \Gamma$ we take any $g\in\aut\Gamma$ such
that $\Delta_\Gamma(g)=\pi$. Consider
\begin{equation}\label{permg}
g x = z,\,g y= w\q \,x \in \el_i,\,y\in\el_j,\,z\in\el_{\pi
(i)},\,w\in\el_{\pi(j)},\,i,j\in I.
\end{equation}
Then for all $x\in\el_i,\,y\in\el_j$ the bilinear mapping ${[x,y]_{\ep^\pi}=\ep_{\pi(i)
\pi (j)} [x,y]}$ and the Lie bracket ${[x,y]_{\ep}=\ep_{ij} [x,y]}$ satisfy
\begin{equation}\label{permiz}
[x,y]_{\ep^\pi}=\ep_{\pi (i) \pi (j)} [x,y]=\ep_{\pi (i) \pi (j)}g^{-1}[z,w]=g^{-1}[g x,g
y]_{\ep}.
\end{equation}
Hence $\el^{\ep^\pi}$ is a Lie algebra and $g$ is an isomorphism between $\el^{\ep^\pi}$
and $\el^\ep$.
\end{proof}

\section{Symmetries of the contraction system}

In other words, lemma \ref{nor2} says that for a given contraction matrix $\ep$ it is
possible to construct new contraction matrices $\ep^\pi,\,\pi \in
\Delta_\Gamma\aut\Gamma$. Of course, the new matrices $\ep^\pi $ have to be the solutions
of the contraction system. We obtained the substitution
$\ep_{ij}\mapsto\ep_{\pi(i)\pi(j)}, \,\pi\in\Delta_\Gamma\aut\Gamma$ under which the set
of solutions of the contraction system is invariant. Now we can also define the {\bf
action of $\Delta_\Gamma\aut\Gamma$ on the contraction system $\es$}: each equation of
$\es$ is labeled by a triple of grading indices and we write $e(i\:j\: k)\in \es$ in the
form
\begin{equation}\label{eqs1}
e(i\:j\:k):\, [x,[y,z]_\ep]_\ep + \mbox{cyclically} =0\q (\forall x \in \el_i)(\forall
y\in \el_j)(\forall z \in \el_k);
\end{equation}
then for each $\pi \in \Delta_\Gamma\aut\Gamma$ we define the action
\begin{equation}\label{akc}
e(i\: j\: k)\,\mapsto\, e(\pi(i)\: \pi(j)\: \pi(k)).
\end{equation}
Note that equation $e(\pi(i)\: \pi(j)\: \pi(k))$ can be written as
\begin{equation}\label{hhh}
e(\pi(i)\:\pi(j)\:\pi(k)):\, [gx,[gy,gz]_\ep]_\ep + \mbox{cyclically} =0\q (\forall x \in
\el_i)(\forall y\in \el_j)(\forall z \in \el_k),
\end{equation}
where $g\in\aut\Gamma,\,\Delta_\Gamma (g)=\pi\in\Delta_\Gamma\aut\Gamma$. According to
(\ref{permiz}) this is equal to
\begin{equation}\label{hhh2}
g[x,[y,z]_{\ep^\pi}]_{\ep^\pi} + \mbox{cyclically} =0\q (\forall x \in \el_i)(\forall
y\in \el_j)(\forall z \in \el_k)
\end{equation}
and (\ref{hhh2}) is satisfied if and only if
\begin{equation}\label{hhh3}
[x,[y,z]_{\ep^\pi}]_{\ep^\pi} + \mbox{cyclically} =0\q (\forall x \in \el_i)(\forall y\in
\el_j)(\forall z \in \el_k).
\end{equation}
The equation (\ref{hhh3}) is precisely the equation (\ref{eqs1}) after the substitution
$\ep_{ij}\mapsto\ep_{\pi(i)\pi(j)}$. In this way we have not only verified the invariance
of the contraction system (up to equivalence of equations), but also have shown the
method of its construction. Having a starting equation one can write a whole orbit of
equations merely by substituting $\ep_{ij}\mapsto\ep_{\pi(i)\pi(j)}$ till all
$\pi\in\Delta_\Gamma\aut\Gamma$ are exhausted. If we denote unordered k-tuple of grading
indices $i_1,i_2,\dots,i_k\in I$ as $i_1\: i_2 \dots i_k$ and define the {\bf action of
$\Delta_\Gamma\aut\Gamma$ on unordered k-tuples} as
\begin{equation}\label{akc0}
i_1 \:i_2 \:\dots i_k\,\mapsto\, \pi(i_1)\: \pi(i_2) \dots \pi(i_k),\q \pi \in
\Delta_\Gamma\aut\Gamma,
\end{equation}
then it is clear that orbits of equations correspond to orbits of unordered triples of
grading indices.

\subsection{Equivalence of solutions}

For the contraction system $\es$ of a graded Lie algebra (\ref{graff}) we denote the set
of all its solutions as $\er (\es)$. By combining lemma \ref{nor1} and lemma \ref{nor2}
it is easy to see that an equivalence relation on the set $\er (\es)$ naturally arises:
two solutions $\ep_1,\ep_2 \in \er(\es)$ are {\bf equivalent}, $\ep_1\sim\ep_2$, if there
exists a normalization matrix $\alpha$ and $\pi \in \Delta_\Gamma \ \text{Aut}\ \Gamma$
such that
\begin{equation}\label{ekkk}
\ep_1=\alpha\bullet\ep_2^\pi.
\end{equation}

The reflexivity of the relation $\sim$ is clear. To check the symmetry and transitivity
we note that if two solutions are equivalent there exist a diagonal mapping $h\in GL(V)$
defined via formula (\ref{diagz}) and an automorphism $g\in\aut\Gamma,\,\Delta_\Gamma
(g)=\pi$ with the property (\ref{permiz}) satisfying
\begin{equation}\label{eqi}
  gh[x,y]_{\ep_1}=gh[x,y]_{\alpha\bullet\ep_2^\pi}=g[hx,hy]_{\ep_2^\pi}=[ghx,ghy]_{\ep_2}
\end{equation}
and {\it vice versa}, i.e. two solution $\ep_1,\ep_2$ are equivalent, $\ep_1\sim\ep_2$,
if and only if
\begin{equation}\label{eqi2}
 gh[x,y]_{\ep_1}=[ghx,ghy]_{\ep_2}
\end{equation}
holds for all $x\in\el_i,\,y\in\el_j,\,i,j\in I$. It is easy to see that
\begin{equation}\label{hg}
h g=g h^\pi
\end{equation}
where $h^\pi$ is the diagonal mapping defined by
\begin{equation}\label{diagzpi}
h^\pi x = a_{\pi(i)} x\q i \in I,\,x_i \in \el_i.
\end{equation}
Modifying (\ref{eqi2}) we have
\begin{equation*}\label{eqi3}
[h^{-1}g^{-1}x,h^{-1}g^{-1}y]_{\ep_1}=h^{-1}g^{-1}[x,y]_{\ep_2}\q(\forall
x\in\el_i)(\forall y\in\el_j).
\end{equation*}
Using the relation $h^{-1}g^{-1}=g^{-1}(h^{\pi^{-1}})^{-1}$ which follows from (\ref{hg})
we have $\ep_2\sim\ep_1$, i.e. we proved the symmetry of the relation $\sim$. The proof
of transitivity can be carried out in a similar manner: if
$\ep_1\sim\ep_2,\,\ep_2\sim\ep_3$, i.e.
$\ep_1=\alpha\bullet\ep_2^\pi,\,\ep_2=\beta\bullet\ep_3^\sigma$, then there exist
diagonal mappings $h,h'$ and automorphisms $g,g'\in\aut\Gamma$ such that
\begin{align*}
 gh[x,y]_{\ep_1}&=[ghx,ghy]_{\ep_2} \\
 g'h'[x,y]_{\ep_2}&=[g'h'x,g'h'y]_{\ep_3} \q (\forall x\in\el_i)(\forall y\in\el_j).
\end{align*}
Hence using (\ref{hg}) we obtain
\begin{equation}\label{trans2}
 g'g(h')^\pi h[x,y]_{\ep_1}=g'h'gh[x,y]_{\ep_1}=[g'h'ghx,g'h'ghy]_{\ep_3}=[ g'g(h')^\pi hx, g'g(h')^\pi
 hy]_{\ep_3}.
\end{equation}
The diagonal mapping $(h')^\pi h$ and the automorphism $g'g$ then imply the equivalence
$\ep_1\sim\ep_3$.

We conclude with
\begin{cor}\label{cor1}
 {\it Graded contractions corresponding to equivalent solutions are isomorphic.}
\end{cor}
\begin{proof}
See (\ref{eqi2}).
\end{proof}

\section{Symmetry group of the Pauli grading}

In our case, when $\Gamma$ is given by (\ref{slg}), the corresponding MAD-group is equal
to $\mbox{Ad}\,\Pi_n$. One can check the fact that $\mbox{Stab}\,\Gamma=\mbox{Ad}\,\Pi_n$
also directly, $$ \text{Ad}_{P} \, X_{rs} = P Q^{r} P^{s} P^{-1} =
    \omega^{r} X_{rs}, \q
\text{Ad}_{Q} \, X_{rs} = Q Q^{r} P^{s} Q^{-1} =
    \omega^{-s} X_{rs}.$$
We introduce a finite matrix group
\begin{equation}\label{HH}
  H_n= \left\{\left( \begin{array}{cc}
               a&b\\
               c&d
              \end{array}
\right)\ \bigg|\  a,b,c,d\in\mathbb{Z}_n,\    ad-bc=\pm 1  \ {\mathrm{mod}}\ n \right\}.
\end{equation}
This group contains the subgroup of matrices with determinant +1 called $SL(2,\Z_n)$. In
\cite{HPP4} an important theorem is proved:
\begin{thm}\label{quo}
The quotient group $\mathcal{N}(\mbox{Ad}\,\Pi_n)/\mbox{Ad}\,\Pi_n$ is isomorphic to the
group $H_n$,
\begin{equation}\label{izomofni}
\mathcal{N}(\mbox{Ad}\,\Pi_n)/\mbox{Ad}\,\Pi_n\simeq H_n .
\end{equation}
\end{thm}
\bigskip
Denoting by $\pi_A$ the permutation corresponding to the matrices $A\in H_n$, the action
of $\pi_A$ on the indices of the grading group $\Z_n \times \Z_n$ is given by
\begin{equation}\label{per5}
  \pi_A\,(i\,\,j)=(i\,\,j)\,A,
\end{equation}
where matrix multiplication modulo $n$ is found on the right hand side.
\begin{cor}
{\it The symmetry group of the Pauli grading of $sl(n,\Com)$ is isomorphic to the matrix
group $H_n$.}
\end{cor}
\bigskip

\chapter{Solution of $\es_3$}
\section{Simplification of $\es_3$}
We have seen that in the case of the grading $\Gamma_4$ given
explicitly by (\ref{zkos}) the symmetry group
$\Delta_{\Gamma}\aut\Gamma$ is isomorphic to $H_3$. The matrix
group $H_3$ has 48 elements and there exist exactly {\it two}
24-point orbits of grading indices triples. We can choose the
triples {\it (01)(02)(10)} and {\it (01)(10)(11)} as
representative elements of these orbits. Moreover, all 24 elements
from each orbit can be obtained by the action of 24 elements from
$SL(2,\Z_3)\subset H_3$ starting from an arbitrary point. Then for
our choice of representative points our system $\es_3$ can be
written elegantly as
\begin{align}
\es_3^a\,:\,\,&\ep_{(02)(10)A}\ep_{(01)(12)A}-\ep_{(01)(10)A}\ep_{(02)(11)A}=0&\label{fff}
\\
\es_3^b\,:\,\,
&\ep_{(10)(11)A}\ep_{(01)(21)A}-\ep_{(01)(11)A}\ep_{(10)(12)A}=0&\q\forall
A \in SL(2,\Z_3)\label{ffff}
\end{align} where we used abbreviation $\ep_{(i j)(k
l)A}:=\ep_{(i j)A,(k l)A}$. Equations 1-24 ({\bf subsystem
$\es_3^a$}) in Table 1 arose from (\ref{fff}) and equations 25-48
({\bf subsystem $\es_3^b$}) from (\ref{ffff}). Table 1 of the
system $\es_3$ clearly shows the correspondence between matrices
from $SL(2,\Z_3)$ and their equations.

Looking closer at the system $\es_3^a$, we observe that adding equations 1 and 3 we
obtain equation 5. Equation 5 is then satisfied automatically whenever 1 and 3 hold and
is redundant in the system. The question whether or not there exist other similar triples
of equations can be answered by the following consideration making use of (\ref{fff}):
equation 1 can be written in the form
\begin{equation}\label{X1}
\ep_{(01)(10)X}\ep_{(02)(11)X}-\ep_{(01)(10)}\ep_{(02)(11)}=0,\q\q
X=\begin{pmatrix}
  1 & 2 \\
  0 & 1
\end{pmatrix}
\end{equation}
This is caused by the fact that $[(02)(10)][(01)(12)]$ and $[(01)(10)][(02)(11)]$ lie in
the same orbit with respect to the action of $SL(2,\Z_3)$ (analogous to \ref{akc0}),
where the pairs of indices in bracket $[\,\,]$ and the pairs of these brackets are
unordered. The equation generated from equation 1 by the matrix $A=X$
\begin{equation}\label{X2}
\ep_{(01)(10)X^2}\ep_{(02)(11)X^2}-\ep_{(01)(10)X}\ep_{(02)(11)X}=0
\end{equation}
is also contained in $\es_3^a$, due to (\ref{fff}). Adding equations (\ref{X1}) and
(\ref{X2}) we have
\begin{equation}\label{X3}
\ep_{(01)(10)X^2}\ep_{(02)(11)X^2}-\ep_{(01)(10)}\ep_{(02)(11)}=0.
\end{equation}
Since $X^3=1$ holds, equation (\ref{X3}) is generated from equation 1 by matrix $A=X^2$.
Therefore we can conclude that the cosets of the left decomposition of the group
$SL(2,\Z_3)$ with respect to the cyclic subgroup $\{1,X,X^2\}$ then generate just the
triples of dependent equations. The number of these cosets according to the Lagrange's
theorem is $24/3=8$. So we obtained 8 equations (one from each coset) which we eliminate
from the system $\es_3^a$. On the other hand, we observe in $\es_3^b$ that the quadruples
of indices $[(10)(11)][(01)(21)]$ and $[(01)(11)][(10)(12)]$ do {\it not} lie in the same
orbit and in this way the equations are not dependent.

The dependent equations are listed below. Symbolically, for the equation numbers the
following holds: 1 + 3 = 5, 2 + 6 = 4, 11 + 17 = 19, 13 + 7 = 22, 15 + 23 = 10, 12 + 20 =
18, 14 + 21 = 8, 9 + 16 = 24, this means precisely that adding equations 1 and 3 equation
5 is obtained etc. The equations 5, 4, 19, 22, 10, 18, 8, 24 then can be chosen as
redundant and eliminated. Thus the number of equations of the system $\es_3$ is reduced
to 40.
\bigskip
\section{The algorithm of evaluation}
Since a straightforward generalization of the method published in \cite{Spanien} turned
out to be impossible for our case, we had to find another way. Of course, we are left
with a laborious case by case analysis. We found a method simplifying this laborious
analysis; it was even possible to compute the parametric solutions without relying on the
computer. The method is based on the fact that under suitable assumptions the system
$\es_3$ can be easily explicitly solved. But after leaving the assumption, that means
putting zero on the position we had assumed non-zero before, the solution is far more
complicated. This can be bypassed by our algorithm which is based on the following
theorem.
\begin{thm}\label{main}
Let $\er (\es)$ be the set of solutions and $\I$ the set of relevant pairs of unordered
indices of the contraction system $\es$ of a graded Lie algebra $\Gamma : \el=\bigoplus
_{i \in I} \el_i$. For any $\qe \subset \er (\es)$ and
$\pe=\{k_1,\,k_2,\dots,k_m\}\subset\I$ we denote
\begin{align*}
\er_0&:=\left\{\ep\in\er(\es)\big|(\forall \ep_1\in\qe)(\ep\nsim\ep_1) \right\}\\
\er_1&:=\left\{ \ep \in \er_0\, \big|\,(\forall k \in \pe )( \ep_k\neq 0 )\right\}.
\end{align*}
Then the solution $\ep\in\er_0$ is non-equivalent to all solutions in $\er_1$ if and only
if
\begin{align}\label{none}
\ep_{\pi_1 (k_1)} \ep_{\pi_1 (k_2)}\cdots \ep_{\pi_1 (k_m)}&=0 \nonumber\\ \vdots \q\q&
\\ \ep_{\pi_n (k_1)} \ep_{\pi_n (k_2)}\cdots \ep_{\pi_n (k_m)}&=0 \nonumber
\end{align}
holds, where $\{\pi_1,\pi_2,\dots,\pi_n\}=\Delta_\Gamma\aut\Gamma$ is the symmetry group
of the grading $\Gamma$.
\end{thm}
\begin{proof}
For any $\ep\in\er_0$ we have (see \ref{ekkk}):
\begin{align}
(\exists\ep_1\in\er_1)(\ep\sim\ep_1)&\Leftrightarrow \,\,(\exists
\ep_1\in\er_1)(\exists\alpha)(\exists\pi\in\
\Delta_\Gamma\aut\Gamma)(\alpha\bullet\ep^\pi=\ep_1)\label{main1}\\
&\Leftrightarrow\,\,(\exists\alpha)(\exists\pi\in\
\Delta_\Gamma\aut\Gamma)(\alpha\bullet\ep^\pi \in \er_0\wedge(\alpha\bullet\ep^\pi)_k\neq
0,\,\, \forall k \in \pe) \label{main2}\\ &\Leftrightarrow\,\,(\exists\pi\in\
\Delta_\Gamma\aut\Gamma)(\forall k \in \pe)((\ep^\pi)_k\neq 0).\label{main3}
\end{align}
The equivalence (\ref{main1}) is direct consequence of the definition (\ref{ekkk}), the
equivalence (\ref{main2}) expresses the trivial fact that
$(\exists\ep_1\in\er_1)(\alpha\bullet\ep^\pi=\ep_1)\Leftrightarrow (\alpha\bullet\ep^\pi
\in \er_1)$. Since $\alpha\bullet\ep^\pi \in \er_0$ is for any $\ep\in\er_0$
automatically fulfilled and $(\alpha\bullet\ep^\pi)_k\neq 0\Leftrightarrow(\ep^\pi)_k\neq
0$, the equivalence (\ref{main3}) follows.

Negating (\ref{main3}) we obtain
\begin{align*}
(\forall\ep_1\in\er_1)(\ep\nsim \ep_1)\Leftrightarrow \,\,(\forall\pi\in\
\Delta_\Gamma\aut\Gamma)(\exists k \in \pe)((\ep^\pi)_k = 0)
\end{align*}
and this is the statement of the theorem.
\end{proof}
We call the system of equations (\ref{none}) corresponding to the sets $\qe \subset \er
(\es)$ and $\pe\subset\I$ a~{\bf non-equivalence system}.

Repeated use of the theorem leads us to the following algorithm for the evaluation of
solutions:
\begin{enumerate}[1.]
\item we set $\qe=\emptyset$ and suppose we have a set of assumptions $\pe^0\subset\I$. Then
$\er_0=\er(\es)$, we explicitly evaluate $$\er^0=\left\{ \ep\in \er(\es)\,\,|\,\,(\forall
k \in \pe^0)(\ep_k \neq 0) \right\} $$ and write the non-equivalence system $\es^0$ of
equations (\ref{none}) corresponding to  $\qe=\emptyset,\,\pe^0$.
\item we set $\qe=\er^0$ and suppose we have the set $\pe^1\subset\I$. Then
$\er_0=\er(\es\cup\es^0)$, we explicitly evaluate $$\er^1=\left\{ \ep\in
\er(\es\cup\es^0)\,\,|\,\,(\forall k \in \pe^1)(\ep_k \neq 0) \right\} $$ and write the
non-equivalence system $\es^1$ corresponding to $\qe=\er^0,\,\pe^1$.
\item we set $\qe=\er^0\cup\er^1$. Then $\er_0=\er(\es\cup\es^0\cup\es^1)$ and we
continue till we have evaluated the whole $\er(\es)$ up to equivalence, i.e. we have such
$\qe$ that the corresponding set $\er_0$ is empty or trivial.
\end{enumerate}
\bigskip

We observe that the algorithm crucially depends on the choice of the assumptions sets
$\pe^0,\,\pe^1,\dots$. Since the system $\es_3$ can be solved explicitly nicely assuming
two of its variables non-zeros, we develop a theory for pairs from $\I$. For fixed $k \in
\I$ we define an equivalence {\bf relation $\equiv^k$} on the set $\I^k:=\I\setminus
\{k\}$: for $i,j\in\I^k$
\begin{equation}\label{ekviv}
i\equiv^k
j\,\,\,\Leftrightarrow\,\,(\exists\pi\in\Delta_\Gamma\aut\Gamma)(\pi\left(i\,\,k\right)
=\left( j\,\,k\right) ),
\end{equation}
where $(i\,\,j)$ denotes an unordered pair of $i,j\in\I$ and naturally $\pi\left(
i\,\,k\right):=\left( \pi(i)\,\,\pi(k)\right)$. The usage of this equivalence will be
seen on our concrete evaluation. We will make use of the following example:
\begin{example}\label{costes} The set of relevant indices $\I_3$ has $24$ elements which
are explicitly written in matrix (\ref{epsi}). We choose the index $k=(01)(10)$ and in
Table 2 we list nine equivalence classes $\I_1^k,\dots,\I_9^k$ of the equivalence
$\equiv^k$: $$\begin{tabular}{c}
\parbox[l][20pt][c]{3pt}{}{\it Table 2: The equivalence classes of $\equiv^{(01)(10)}$} \\
\begin{tabular}[b]{c|l}\hline \parbox[l][20pt][c]{3pt}{}$ \I_1^k$ & (11)(12), (11)(21),
(22)(12), (22)(21) \\ \hline
\parbox[l][20pt][c]{3pt}{}$ \I_2^k$ & (01)(11), (10)(11), (01)(12),
(10)(21) \\ \hline \parbox[l][20pt][c]{3pt}{}$\I_3^k$ & (02)(22), (20)(22), (02)(21),
(20)(12) \\ \hline
\parbox[l][20pt][c]{3pt}{}$\I_4^k$ & (01)(20), (02)(10) \\ \hline
\parbox[l][20pt][c]{3pt}{}$\I_5^k$ & (01)(22), (10)(22) \\ \hline \parbox[l][20pt][c]{3pt}{}$\I_6^k$ & (01)(21),
(10)(12)
\\ \hline \parbox[l][20pt][c]{3pt}{}$\I_7^k$ & (02)(11), (20)(11) \\ \hline \parbox[l][20pt][c]{3pt}{}$\I_8^k$ & (02)(12), (20)(21)
\\ \hline
\parbox[l][20pt][c]{3pt}{}$\I_9^k$ & (02)(20) \\ \hline
\end{tabular}
\end{tabular}$$
\end{example}

\section{Evaluation of $\es_3$ solutions}
\begin{align*}1.\q\q&\er^0=\left\{ \ep\in \er(\es_3)\,\,|\,\,\ep_{(01)(10)}\neq
0,\,\ep_{(22)(21)}\neq 0 \right\}\\ &\es^0:\,\,\ep_{(01)(10)A}\ep_{(22)(21)A}=0\q\forall
A\in H_3\\ 2.\q\q &\er^1=\left\{ \ep\in \er(\es_3\cup\es^0)\,\,|\,\,\ep_{(01)(10)}\neq
0,\,\ep_{(10)(11)}\neq 0,\,\ep_{(01)(22)}\neq 0 \right\}\\
&\es^1:\,\,\ep_{(01)(10)A}\ep_{(10)(11)A}\ep_{(01)(22)A}=0\q\forall A\in H_3\\ 3.\q\q
&\er^2=\left\{ \ep\in \er(\es_3\cup\es^0\cup\es^1)\,\,|\,\,\ep_{(01)(10)}\neq
0,\,\ep_{(10)(11)}\neq 0 \right\}\\ &\es^2:\,\,\ep_{(01)(10)A}\ep_{(10)(11)A}=0\q\forall
A\in H_3\\  4.\q\q &\er^3=\left\{ \ep\in
\er(\es_3\cup\es^0\cup\es^1\cup\es^2)\,\,|\,\,\ep_{(01)(10)}\neq 0,\,\ep_{(02)(22)}\neq 0
\right\}\\ &\es^3:\,\,\ep_{(01)(10)A}\ep_{(02)(22)A}=0\q\forall A\in H_3\\ 5.\q\q
&\er^4=\left\{ \ep\in
\er(\es_3\cup\es^0\cup\es^1\cup\es^2\cup\es^3)\,\,|\,\,\ep_{(01)(10)}\neq 0 \right\}\\
&\es^4:\,\,\ep_{(01)(10)A}=0\q\forall A\in H_3
\end{align*}

\begin{enumerate}[\text{ad} 1.]
\item We present the evaluation of $\er^0$ in detail. First note that equation 37 implies $\ep_{(01)(22)}  \neq 0,\, \ep_{(20)(21)} \neq 0$. Now we solve explicitly.
The equation number from which each equality follows is listed before the colon. The
already evaluated unknowns are substituted into each equation.

\begin{tabular}[l]{cl}
$\parbox[l][22pt][c]{3pt}{}37$&$:\q \ep_{(22)(21)} = \frac{\ep_{(01)(22)}
\ep_{(20)(21)}}{\ep_{(01)(10)}}$\\ $\parbox[l][22pt][c]{3pt}{}35$&$:\q \ep_{(01)(11)} =
\frac{\ep_{(01)(21)} \ep_{(20)(22)}}{\ep_{(20)(21)}}$\\
$\parbox[l][22pt][c]{3pt}{}44$&$:\q \ep_{(22)(12)} = \frac{\ep_{(10)(22)}
\ep_{(02)(12)}}{\ep_{(01)(10)}}$\\ $\parbox[l][22pt][c]{3pt}{}31$&$:\q \ep_{(11)(21)} =
\frac{\ep_{(01)(21)} \ep_{(10)(22)}}{\ep_{(01)(10)}}$\\
$\parbox[l][22pt][c]{3pt}{}47$&$:\q \ep_{(20)(12)} = \frac{\ep_{(01)(12)}
\ep_{(10)(22)}}{ \ep_{(01)(22)} }$\\ $\parbox[l][22pt][c]{3pt}{}26$&$:\q \ep_{(02)(22)} =
\frac{\ep_{(20)(22)} \ep_{(02)(12)}}{\ep_{(20)(21)}}$\\
$\parbox[l][22pt][c]{3pt}{}3$&$:\q \ep_{(02)(11)}= \frac{ \ep_{(01)(11)} \ep_{(02)(12)}}{
\ep_{(01)(10)} }=\frac{\ep_{(01)(21)} \ep_{(20)(22)} \ep_{(02)(12)} }{\ep_{(20)(21)}
\ep_{(01)(10)}}$\\ $\parbox[l][22pt][c]{3pt}{}21$&$:\q \ep_{(11)(12)} =
\frac{\ep_{(02)(12)} \ep_{(11)(21)} }{\ep_{(20)(21)} }=\frac{\ep_{(02)(12)}
\ep_{(01)(21)} \ep_{(10)(22)}}{\ep_{(20)(21)} \ep_{(01)(10)}}$\\
$\parbox[l][22pt][c]{3pt}{}29$&$:\q \ep_{(10)(12)} = \frac{ \ep_{(01)(10)} \ep_{(11)(12)}
}{\ep_{(01)(22)}  }= \frac{\ep_{(01)(10)} \ep_{(02)(12)} \ep_{(01)(21)} \ep_{(10)(22)}
}{\ep_{(01)(22)} \ep_{(01)(10)} \ep_{(20)(21)}}=\frac{\ep_{(10)(22)} \ep_{(02)(12)}
\ep_{(01)(21)}}{\ep_{(20)(21)} \ep_{(01)(22)}}$\\ $\parbox[l][22pt][c]{3pt}{}39$&$:\q
\ep_{(01)(20)} = \frac{\ep_{(20)(22)} \ep_{(01)(12)} \ep_{(01)(10)} }{\ep_{(01)(22)}
\ep_{(20)(21)} }$\\ $\parbox[l][22pt][c]{3pt}{}15$&$:\q \ep_{(20)(11)} =
\frac{\ep_{(01)(11)} \ep_{(22)(12)} }{\ep_{(01)(22)} }=\frac{\ep_{(01)(21)}
\ep_{(20)(22)} \ep_{(10)(22)} \ep_{(02)(12)}}{\ep_{(20)(21)} \ep_{(01)(10)}
\ep_{(01)(22)} }$\\ $\parbox[l][22pt][c]{3pt}{}17$&$:\q \ep_{(10)(11)} =
\frac{\ep_{(01)(10)} \ep_{(20)(11)} }{\ep_{(20)(21)} }=\frac{\ep_{(20)(22)}
\ep_{(10)(22)} \ep_{(02)(12)} \ep_{(01)(21)}}{ \ep_{(01)(22)} \ep_{(20)(21)}^2}$\\
$\parbox[l][22pt][c]{3pt}{}30$&$:\q \ep_{(02)(20)} = \frac{ \ep_{(20)(21)} \ep_{(02)(11)}
}{\ep_{(22)(21)} }= \frac{\ep_{(01)(21)} \ep_{(20)(22)} \ep_{(02)(12)}}{\ep_{(20)(21)}
\ep_{(01)(22)}}$\\ $\parbox[l][22pt][c]{3pt}{}28$&$:\q \ep_{(02)(10)} =
\frac{\ep_{(02)(21)} \ep_{(20)(22)} \ep_{(01)(10)} }{\ep_{(01)(22)} \ep_{(20)(21)}}$\\
$\parbox[l][22pt][c]{3pt}{}46$&$:\q \ep_{(10)(21)} = \frac{ \ep_{(10)(22)} \ep_{(02)(21)}
}{ \ep_{(01)(22)}}$\\
\end{tabular}

Up to now we have used 15 out of 40 equations. But at this moment one can check that the
rest of equations is either identically fulfilled or their fulfillment is a consequence
of two equations, 1 and 13. After the substitution of already evaluated unknowns they are
obtained in the form

\begin{tabular}[l]{cl}
$\parbox[l][22pt][c]{3pt}{}1$&$:\q \ep_{(20)(22)} \ep_{(01)(10)} \ep_{(01)(12)}
\ep_{(02)(21)} = \ep_{(20)(22)} \ep_{(01)(22)} \ep_{(01)(21)} \ep_{(02)(12)} $\\
$\parbox[l][22pt][c]{3pt}{}13$&$:\q \ep_{(10)(22)} \ep_{(01)(10)} \ep_{(01)(12)}
\ep_{(02)(21)} = \ep_{(10)(22)} \ep_{(01)(22)} \ep_{(01)(21)} \ep_{(02)(12)}. $
\end{tabular}

The discussion of these two equations is easy and here we state the result:
$$\er^0=\{\ep^0_1,\ep^0_2,\ep^0_3,\ep^0_4\},$$ where

\begin{tabular}[l]{l}
$\ep^0_1 =  \left( {\begin{smallmatrix} 0 & 0 & {t_{1}} & 0 & 0 & {t_{4}} & {x_{5}} &
{x_{6}} \\ 0 & 0 & 0 & 0 & 0 & 0 & {x_{11}} & {x_{12}} \\ {t_{1}} & 0 & 0 & 0 & 0 & 0 & 0
& 0 \\ 0 & 0 & 0 & 0 & 0 & 0 & 0 & {t_{20}} \\ 0 & 0 & 0 & 0 & 0 & 0 & 0 & 0 \\ {t_{4}} &
0 & 0 & 0 & 0 & 0 & 0 & { \frac {{t_{4}}\, {t_{20}}}{{t_{1}}}}  \\  {x_{5}} & {x_{11}} &
0 & 0 & 0 & 0 & 0 & 0 \\ {x_{6}} & {x_{12}} & 0 & {t_{20}} & 0 & { \frac {{t
_{4}}\,{t_{20}}}{{t_{1}}}}  & 0 & 0
\end{smallmatrix}}
 \right)$ \\
$\ep^0_2 =  \left( {\begin{smallmatrix} 0& \,0& {t_{1}}& { \frac {{x_{18}}{t_{5}
}{t_{1}}}{{t_{4}}{t_{20}}}} & { \frac {{x_{6 }}{x_{18}}}{{t_{20}}}} & {t_{4}}& {t_{5}}&
{x_{6}}
 \\
0& 0& { \frac {{x_{11}}{x_{6}}{x_{18}}}{ {t_{5}}{t_{20}}}} & { \frac {{x_{11}}{x_{6}}
{x_{18}}}{{t_{20}}{t_{4}}}} & { \frac {{x_{ 11}}{x_{6}}{x_{18}}}{{t_{20}}{t_{1}}}} &  {
\frac {{x_{18}}{x_{11}}}{{t_{20}}}} & {x_{11 }}& { \frac {{x_{11}}{x_{6}}{t_{4}}}{{t_{5}}
{t_{1}}}}  \\ {t_{1}}& { \frac {{x_{11}}{x_{6}}{x_{18}}}{{ t_{5}}{t_{20}}}} & 0& 0& {
\frac {{x_{ 18}}{x_{14}}{x_{11}}{x_{6}}}{{t_{4}}{t_{20}}^{2}}} &  {x_{14}}& { \frac
{{x_{14}}{x_{11}}{x_{6}}}{ {t_{20}}{t_{4}}}} & { \frac {{x_{14}}{x_{11}
}{x_{6}}}{{t_{5}}{t_{1}}}}  \\  { \frac {{x_{18}}{t_{5}}{t_{1}}}{{t_{4}}{t_{20 }}}} & {
\frac {{x_{11}}{x_{6}}{x_{18}}}{{t _{20}}{t_{4}}}} & 0& 0& { \frac {{x_{
18}}{x_{14}}{x_{11}}{x_{6}}}{{t_{1}}{t_{4}}{t_{20}}}} & {x_{18}}& { \frac
{{t_{5}}{x_{14}}}{{t_{ 4}}}} & {t_{20}} \\  { \frac {{x_{6}}{x_{18}}}{{t_{20}}}} & {
\frac {{x_{11}}{x_{6}}{x_{18}}}{{t_{20}}{t_{ 1}}}} & { \frac {{x_{18}}{x_{14}}{x_{11}}{x
_{6}}}{{t_{4}}{t_{20}}^{2}}} & { \frac {{x_{18}}
{x_{14}}{x_{11}}{x_{6}}}{{t_{1}}{t_{4}}{t_{20}}}} & 0 & 0& { \frac
{{x_{14}}{x_{11}}{x_{6}}}{{t_{1 }}{t_{20}}}} & { \frac {{x_{6}}{x_{14}}}{{t_{1 }}}}
\\  {t_{4}}& { \frac {{x_{18}}{x_{11}}}{{t_{20}}} } & {x_{14}}& {x_{18}}& 0&
0& { \frac {{x_{14}}{x_{11}}}{{t_{1}}}} & { \frac { {t_{4}}{t_{20}}}{{t_{1}}}}
\\  {t_{5}}& {x_{11}}& { \frac {{x_{14}}{x_{11
}}{x_{6}}}{{t_{20}}{t_{4}}}} & { \frac {{t_{ 5}}{x_{14}}}{{t_{4}}}} & { \frac {{x_{14}}{x
_{11}}{x_{6}}}{{t_{1}}{t_{20}}}} & { \frac { {x_{14}}{x_{11}}}{{t_{1}}}} & 0& 0
\\  {x_{6}}& { \frac {{x_{11}}{x_{6}}{t_{4}}}{{t _{5}}{t_{1}}}} & { \frac
{{x_{14}}{x_{11}} {x_{6}}}{{t_{5}}{t_{1}}}} & {t_{20}}& { \frac
{{x_{6}}{x_{14}}}{{t_{1}}}} & { \frac {{ t_{4}}{t_{20}}}{{t_{1}}}} & 0& 0
\end{smallmatrix}}
 \right)$\\
$\ep^0_3 =  \left( {\begin{smallmatrix} 0 & 0 & {t_{1}} & 0 & 0 & {t_{4}} & 0 & 0
\\ 0 & 0 & { \frac {{x_{12}}{x_{18}}{t_{1}}}{{t_{4} }{t_{20}}}}  & 0 & 0 &
{ \frac {{x_{18}}{x_{11}} }{{t_{20}}}}  & {x_{11}} & {x_{12}} \\  {t_{1}} & { \frac
{{x_{12}}{x_{18}}{t_{1}}}{{t_{ 4}}{t_{20}}}}  & 0 & 0 & 0 & {x_{14}} & 0 & { \frac
{{x_{14}}{x_{12}}}{{t_{4}}}}  \\  0 & 0 & 0 & 0 & 0 & {x_{18}} & 0 & {t_{20}} \\ 0 & 0 &
0 & 0 & 0 & 0 & 0 & 0 \\ {t_{4}} & { \frac {{x_{18}}{x_{11}}}{{t_{20}}}}
 & {x_{14}} & {x_{18}} & 0 & 0 & { \frac {{x_{14}}
{x_{11}}}{{t_{1}}}}  & { \frac {{t_{4}}{t_{20}}}{{ t_{1}}}}  \\  0 & {x_{11}} & 0 & 0 & 0
& { \frac {{x_{14}}{x_{11 }}}{{t_{1}}}}  & 0 & 0 \\  0 & {x_{12}} & { \frac
{{x_{14}}{x_{12}}}{{t_{4}}} }  & {t_{20}} & 0 & { \frac {{t_{4}}{t_{20}}}{{t_{ 1}}}} & 0
& 0
\end{smallmatrix}}
 \right)$ \\
$\ep^0_4 =  \left( {\begin{smallmatrix} 0 & 0 & {t_{1}} & 0 & { \frac
{{x_{6}}{x_{18}}}{{t _{20}}}}  & {t_{4}} & 0 & {x_{6}} \\  0 & 0 & { \frac
{{x_{12}}{x_{18}}{t_{1}}}{{t_{4} }{t_{20}}}}  & 0 & 0 & 0 & 0 & {x_{12}} \\ {t_{1}} & {
\frac {{x_{12}}{x_{18}}{t_{1}}}{{t_{ 4}}{t_{20}}}} & 0 & 0 & 0 & {x_{14}} & 0 & { \frac
{{x_{14}}{x_{12}}}{{t_{4}}}}  \\  0 & 0 & 0 & 0 & 0 & {x_{18}} & 0 & {t_{20}}
\\ { \frac {{x_{6}}{x_{18}}}{{t_{20}}}}  & 0 & 0 & 0
 & 0 & 0 & 0 & { \frac {{x_{6}}{x_{14}}}{{t_{1}}}
}  \\  {t_{4}} & 0 & {x_{14}} & {x_{18}} & 0 & 0 & 0 & { \frac
{{t_{4}}{t_{20}}}{{t_{1}}}}  \\  0 & 0 & 0 & 0 & 0 & 0 & 0 & 0 \\ {x_{6}} & {x_{12}} & {
\frac {{x_{14}}{x_{12}}}{{t _{4}}}}  & {t_{20}} & { \frac {{x_{6}}{x_{14}}}{{t _{1}}}} &
{ \frac {{t_{4}}{t_{20}}}{{t_{1}}}}  & 0 & 0
\end{smallmatrix}}
 \right).$
\end{tabular}

In the notation of explicit matrices we use a convention that non-zero parameters are
denoted by the letter $t$ and others which can be zero by the letter $x$. Henceforth the
same convention will be used.
\item Note that the system of 48 equations $\es^0$ together with $\ep_{(01)(10)}\neq 0$ enforces zeros
on all positions from $\I_1^k$. {\it In general we can say that after solving with the
assumption $\ep_k\neq 0,\,\ep_i\neq 0$ the corresponding non-equivalence system and the
assumption $\ep_k\neq 0$ will enforce zeros on all positions $j,\,j\equiv^k i$.} That is
exactly the reason why we chose $(22)(21)\in\I_1^k,\,(10)(11)\in\I_2^k,\,(02)(22)\in
\I_3^k$ for the evaluation of the sets $\er^0,\,\er^1$ and $\er^2,\,\er^3$ respectively.
Moreover, the assumption $\ep_{(10)(11)}\neq 0$ and $\es^0$ enforces further 4 zeros.
Since the assumption $\ep_{(01)(10)}\neq 0,\,\ep_{(10)(11)}\neq 0,\,\ep_{(01)(22)}\neq 0$
gives us in some way a special solution we put it in a single set: $$\er^1=\{\ep^1\}$$
$$\ep^1 = \left( {\begin{smallmatrix} 0 & 0 & {t_{1}} & {x_{2}} & {x_{3}} & {t_{4}} & 0 &
0 \\ 0 & 0 & {x_{7}} & 0 & 0 & {x_{10}} & 0 & 0 \\ {t_{1}} & {x_{7}} & 0 & 0 & {t_{13}} &
{x_{14}} & 0 & 0 \\ {x_{2}} & 0 & 0 & 0 & 0 & {x_{18}} & 0 & 0 \\ {x_{3}} & 0 & {t_{13}}
& 0 & 0 & 0 & 0 & 0 \\ {t_{4}} & {x_{10}} & {x_{14}} & {x_{18}} & 0 & 0 & 0 & 0
\\ 0 & 0 & 0 & 0 & 0 & 0 & 0 & 0 \\ 0 & 0 & 0 & 0 & 0 & 0 & 0 & 0
\end{smallmatrix}}
 \right)$$
\item The solutions with assumption $\ep_{(01)(10)}\neq 0,\,\ep_{(10)(11)}\neq 0$
non-equivalent to those in $\er^1$ and $\er^0$ are listed below:
$$\er^2=\{\ep^2_1,\ep^2_2,\ep^2_3,\ep^2_4,\ep^2_5,\ep^2_6,\ep^2_7,\ep^2_8\}$$
\end{enumerate}
$$\ep^2_1 =  \left( {\begin{smallmatrix} 0 & 0 & {t_{1}} & {x_{2}} & 0 & 0 & 0 & 0
\\ 0 & 0 & 0 & 0 & 0 & 0 & 0 & 0 \\ {t_{1}} & 0 & 0 & 0 & {t_{13}} & {x_{14}} & 0 &
{x_{16}} \\ {x_{2}} & 0 & 0 & 0 & { \frac {{x_{2}}\,{x_{16}}}{{t _{1}}}}  & {x_{18}} &
{x_{19}} & { \frac {{x_{2}}\,{ x_{16}}}{{t_{13}}}}  \\  0 & 0 & {t_{13}} & { \frac
{{x_{2}}\,{x_{16}}}{{t_{1 }}}}  & 0 & 0 & 0 & 0 \\  0 & 0 & {x_{14}} & {x_{18}} & 0 & 0 &
0 & 0 \\ 0 & 0 & 0 & {x_{19}} & 0 & 0 & 0 & 0 \\ 0 & 0 & {x_{16}} & { \frac
{{x_{2}}\,{x_{16}}}{{t_{ 13}}}}  & 0 & 0 & 0 & 0
\end{smallmatrix}}
 \right)
\ep^2_2 =  \left( {\begin{smallmatrix} 0 & 0 & {t_{1}} & {x_{2}} & 0 & 0 & 0 & 0 \\ 0 & 0
& 0 & 0 & 0 & 0 & 0 & 0 \\ {t_{1}} & 0 & 0 & 0 & {t_{13}} & {x_{14}} & {t_{15}} &
{x_{16}}
 \\
{x_{2}} & 0 & 0 & 0 & { \frac {{x_{2}}\,{x_{16}}}{{t _{1}}}}  & 0 & {x_{19}} & { \frac
{{x_{2}}\,{x_{16}} }{{t_{13}}}}  \\  0 & 0 & {t_{13}} & { \frac {{x_{2}}\,{x_{16}}}{{t_{1
}}}}  & 0 & 0 & 0 & 0 \\  0 & 0 & {x_{14}} & 0 & 0 & 0 & 0 & 0 \\ 0 & 0 & {t_{15}} &
{x_{19}} & 0 & 0 & 0 & 0 \\ 0 & 0 & {x_{16}} & { \frac {{x_{2}}\,{x_{16}}}{{t_{ 13}}}}  &
0 & 0 & 0 & 0
\end{smallmatrix}}
 \right)$$ $$
\ep^2_3 =  \left( {\begin{smallmatrix} 0 & 0 & {t_{1}} & {x_{2}} & 0 & 0 & 0 & 0
\\ 0 & 0 & {t_{7}} & 0 & 0 & 0 & 0 & 0 \\ {t_{1}} & {t_{7}} & 0 & 0 & {t_{13}} & {x_{14}}
& 0 & {x_{16}} \\ {x_{2}} & 0 & 0 & 0 & { \frac {{x_{2}}\,{x_{16}}}{{t _{1}}}}  &
{x_{18}} & 0 & { \frac {{x_{2}}\,{x_{16}} }{{t_{13}}}}  \\  0 & 0 & {t_{13}} & { \frac
{{x_{2}}\,{x_{16}}}{{t_{1 }}}}  & 0 & 0 & 0 & 0 \\  0 & 0 & {x_{14}} & {x_{18}} & 0 & 0 &
0 & 0 \\ 0 & 0 & 0 & 0 & 0 & 0 & 0 & 0 \\ 0 & 0 & {x_{16}} & { \frac
{{x_{2}}\,{x_{16}}}{{t_{ 13}}}}  & 0 & 0 & 0 & 0
\end{smallmatrix}}
 \right)
\ep^2_4 =  \left( {\begin{smallmatrix} 0 & 0 & {t_{1}} & {x_{2}} & 0 & 0 & 0 & 0
\\ 0 & 0 & {t_{7}} & 0 & 0 & 0 & 0 & 0 \\ {t_{1}} & {t_{7}} & 0 & 0 & {t_{13}} & {x_{14}}
& {t_{15}} & {x_{ 16}} \\ {x_{2}} & 0 & 0 & 0 & { \frac {{x_{2}}\,{x_{16}}}{{t _{1}}}}  &
0 & 0 & { \frac {{x_{2}}\,{x_{16}}}{{t_{ 13}}}}  \\  0 & 0 & {t_{13}} & { \frac
{{x_{2}}\,{x_{16}}}{{t_{1 }}}}  & 0 & 0 & 0 & 0 \\  0 & 0 & {x_{14}} & 0 & 0 & 0 & 0 & 0
\\ 0 & 0 & {t_{15}} & 0 & 0 & 0 & 0 & 0 \\ 0 & 0 & {x_{16}} & { \frac
{{x_{2}}\,{x_{16}}}{{t_{ 13}}}}  & 0 & 0 & 0 & 0
\end{smallmatrix}}
 \right) $$ $$
\ep^2_5 =  \left( {\begin{smallmatrix} 0 & 0 & {t_{1}} & {x_{2}} & 0 & 0 & 0 & 0
\\ 0 & 0 & {t_{7}} & {t_{8}} & 0 & 0 & 0 & 0 \\ {t_{1}} & {t_{7}} & 0 & 0 & {t_{13}} & {
\frac {{x_{ 15}}\,{x_{18}}}{{t_{8}}}}  & {x_{15}} & {x_{16}} \\  {x_{2}} & {t_{8}} & 0 &
0 & { \frac {{x_{2}}\,{x_{16 }}}{{t_{1}}}}  & {x_{18}} & { \frac {{x_{15}}\,{x_{
18}}}{{t_{7}}}}  & { \frac {{x_{2}}\,{x_{16}}}{{t_{ 13}}}}
\\  0 & 0 & {t_{13}} & { \frac {{x_{2}}\,{x_{16}}}{{t_{1 }}}} & 0 & 0 &
0 & 0 \\  0 & 0 & { \frac {{x_{15}}\,{x_{18}}}{{t_{8}}}}  & {x _{18}} & 0 & 0 & 0 & 0 \\
 0 & 0 & {x_{15}} & { \frac {{x_{15}}\,{x_{18}}}{{t_{ 7}}}}  & 0 & 0 & 0 & 0 \\
 0 & 0 & {x_{16}} & { \frac {{x_{2}}\,{x_{16}}}{{t_{ 13}}}}  & 0 & 0 & 0 & 0
\end{smallmatrix}}
 \right)
\ep^2_6 =  \left( {\begin{smallmatrix} 0 & 0 & {t_{1}} & {x_{2}} & 0 & 0 & 0 & 0
\\ 0 & 0 & 0 & {t_{8}} & 0 & 0 & 0 & 0 \\ {t_{1}} & 0 & 0 & 0 & {t_{13}} & 0 & 0 &
{x_{16}} \\ {x_{2}} & {t_{8}} & 0 & 0 & { \frac {{x_{2}}\,{x_{16 }}}{{t_{1}}}}  &
{x_{18}} & {x_{19}} & { \frac {{x_{ 2}}\,{x_{16}}}{{t_{13}}}}  \\  0 & 0 & {t_{13}} & {
\frac {{x_{2}}\,{x_{16}}}{{t_{1 }}}}  & 0 & 0 & 0 & 0 \\  0 & 0 & 0 & {x_{18}} & 0 & 0 &
0 & 0 \\ 0 & 0 & 0 & {x_{19}} & 0 & 0 & 0 & 0 \\ 0 & 0 & {x_{16}} & { \frac
{{x_{2}}\,{x_{16}}}{{t_{ 13}}}}  & 0 & 0 & 0 & 0
\end{smallmatrix}}
 \right)$$ $$
\ep^2_7 =  \left( {\begin{smallmatrix} 0 & 0 & {t_{1}} & {x_{2}} & 0 & 0 & 0 & 0
\\ 0 & 0 & 0 & {t_{8}} & 0 & 0 & 0 & 0 \\ {t_{1}} & 0 & 0 & 0 & {t_{13}} & 0 & {t_{15}} &
{x_{16}} \\ {x_{2}} & {t_{8}} & 0 & 0 & { \frac {{x_{2}}\,{x_{16 }}}{{t_{1}}}}  & 0 &
{x_{19}} & { \frac {{x_{2}}\,{x _{16}}}{{t_{13}}}}  \\  0 & 0 & {t_{13}} & { \frac
{{x_{2}}\,{x_{16}}}{{t_{1 }}}}  & 0 & 0 & 0 & 0 \\  0 & 0 & 0 & 0 & 0 & 0 & 0 & 0 \\ 0 &
0 & {t_{15}} & {x_{19}} & 0 & 0 & 0 & 0 \\ 0 & 0 & {x_{16}} & { \frac
{{x_{2}}\,{x_{16}}}{{t_{ 13}}}}  & 0 & 0 & 0 & 0
\end{smallmatrix}}
 \right)
\ep^2_8 =  \left( {\begin{smallmatrix} 0 & 0 & {t_{1}} & {x_{2}} & {x_{3}} & 0 & 0 & 0 \\
0 & 0 & {x_{7}} & {x_{8}} & 0 & {x_{10}} & 0 & 0 \\ {t_{1}} & {x_{7}} & 0 & 0 & {t_{13}}
& 0 & 0 & 0 \\ {x_{2}} & {x_{8}} & 0 & 0 & 0 & {x_{18}} & 0 & 0 \\ {x_{3}} & 0 & {t_{13}}
& 0 & 0 & 0 & 0 & 0 \\ 0 & {x_{10}} & 0 & {x_{18}} & 0 & 0 & 0 & 0 \\ 0 & 0 & 0 & 0 & 0 &
0 & 0 & 0 \\ 0 & 0 & 0 & 0 & 0 & 0 & 0 & 0
\end{smallmatrix}}
 \right)$$

\begin{enumerate}[ad 4.]
\item  Now we can of course ignore the equations $\es^1$ because they are satisfied identically due to the system
$\es^2$. We list the next set $$\er^3=\{\ep^3_1,\ep^3_2,\ep^3_3,\ep^3_4\}$$ $$\ep^3_1 =
\left( {\begin{smallmatrix} 0 & 0 & {t_{1}} & {x_{2}} & 0 & 0 & 0 & 0 \\ 0 & 0 & 0 & 0 &
0 & {t_{10}} & {t_{11}} & 0 \\ {t_{1}} & 0 & 0 & 0 & 0 & 0 & 0 & 0 \\ {x_{2}} & 0 & 0 & 0
& 0 & 0 & 0 & 0 \\ 0 & 0 & 0 & 0 & 0 & 0 & 0 & 0 \\ 0 & {t_{10}} & 0 & 0 & 0 & 0 & 0 & 0
\\ 0 & {t_{11}} & 0 & 0 & 0 & 0 & 0 & 0 \\ 0 & 0 & 0 & 0 & 0 & 0 & 0 & 0
\end{smallmatrix}}
 \right)
\ep^3_2 =  \left( {\begin{smallmatrix} 0 & 0 & {t_{1}} & {x_{2}} & 0 & 0 & 0 & 0 \\ 0 & 0
& {x_{7}} & 0 & 0 & {t_{10}} & 0 & 0 \\ {t_{1}} & {x_{7}} & 0 & 0 & 0 & 0 & 0 & 0 \\
{x_{2}} & 0 & 0 & 0 & 0 & {x_{18}} & 0 & 0 \\ 0 & 0 & 0 & 0 & 0 & 0 & 0 & 0 \\ 0 &
{t_{10}} & 0 & {x_{18}} & 0 & 0 & 0 & 0 \\ 0 & 0 & 0 & 0 & 0 & 0 & 0 & 0 \\ 0 & 0 & 0 & 0
& 0 & 0 & 0 & 0
\end{smallmatrix}}
 \right)$$ $$
\ep^3_3 =  \left( {\begin{smallmatrix} 0 & 0 & {t_{1}} & 0 & 0 & {t_{4}} & 0 & 0 \\ 0 & 0
& 0 & 0 & 0 & {t_{10}} & {t_{11}} & 0 \\ {t_{1}} & 0 & 0 & 0 & 0 & 0 & 0 & 0 \\ 0 & 0 & 0
& 0 & 0 & 0 & 0 & 0 \\ 0 & 0 & 0 & 0 & 0 & 0 & 0 & 0 \\ {t_{4}} & {t_{10}} & 0 & 0 & 0 &
0 & 0 & 0 \\ 0 & {t_{11}} & 0 & 0 & 0 & 0 & 0 & 0 \\ 0 & 0 & 0 & 0 & 0 & 0 & 0 & 0
\end{smallmatrix}}
 \right)
\ep^3_4 =  \left( {\begin{smallmatrix} 0 & 0 & {t_{1}} & 0 & 0 & {t_{4}} & 0 & 0 \\ 0 & 0
& {x_{7}} & 0 & 0 & {t_{10}} & 0 & 0 \\ {t_{1}} & {x_{7}} & 0 & 0 & 0 & 0 & 0 & 0 \\ 0 &
0 & 0 & 0 & 0 & 0 & 0 & 0 \\ 0 & 0 & 0 & 0 & 0 & 0 & 0 & 0 \\ {t_{4}} & {t_{10}} & 0 & 0
& 0 & 0 & 0 & 0 \\ 0 & 0 & 0 & 0 & 0 & 0 & 0 & 0 \\ 0 & 0 & 0 & 0 & 0 & 0 & 0 & 0
\end{smallmatrix}}
 \right)$$
\end{enumerate} \begin{enumerate}[ad 5.]\item The systems $\es^0,\,\es^2,\,\es^3$ together
with $\ep_{(01)(10)}\neq 0$ give us 12 zeros and further 20 non-trivial conditions.
Adding two zeros following from $\es_3$ we have: $$\er^4=\{\ep^4_1,\ep^4_2,\ep^4_3\}$$
$$\ep^4_1 =  \left( {\begin{smallmatrix} 0 & 0 & {t_{1}} & 0 & 0 & 0 & 0 & {x_{6}} \\ 0 &
0 & 0 & {x_{8}} & 0 & 0 & {x_{11}} & 0 \\ {t_{1}} & 0 & 0 & 0 & 0 & 0 & {x_{15}} & 0 \\ 0
& {x_{8}} & 0 & 0 & 0 & 0 & 0 & {x_{20}}
\\ 0 & 0 & 0 & 0 & 0 & 0 & 0 & 0 \\ 0 & 0 & 0 & 0 & 0 & 0 & 0 & 0 \\ 0 & {x_{11}} &
{x_{15}} & 0 & 0 & 0 & 0 & 0 \\ {x_{6}} & 0 & 0 & {x_{20}} & 0 & 0 & 0 & 0
\end{smallmatrix}}
 \right)
\ep^4_2 =  \left( {\begin{smallmatrix} 0 & 0 & {t_{1}} & {x_{2}} & 0 & 0 & 0 & 0 \\ 0 & 0
& {x_{7}} & {x_{8}} & 0 & 0 & 0 & 0 \\ {t_{1}} & {x_{7}} & 0 & 0 & 0 & 0 & 0 & 0 \\
{x_{2}} & {x_{8}} & 0 & 0 & 0 & 0 & 0 & 0 \\ 0 & 0 & 0 & 0 & 0 & 0 & 0 & 0 \\ 0 & 0 & 0 &
0 & 0 & 0 & 0 & 0 \\ 0 & 0 & 0 & 0 & 0 & 0 & 0 & 0 \\ 0 & 0 & 0 & 0 & 0 & 0 & 0 & 0
\end{smallmatrix}}
 \right)$$
$$\ep^4_3 =  \left( {\begin{smallmatrix} 0 & 0 & {t_{1}} & 0 & 0 & {x_{4}} & 0 & 0 \\ 0 &
0 & 0 & 0 & 0 & 0 & 0 & 0 \\ {t_{1}} & 0 & 0 & 0 & 0 & {x_{14}} & 0 & 0 \\ 0 & 0 & 0 & 0
& 0 & 0 & 0 & 0 \\ 0 & 0 & 0 & 0 & 0 & 0 & 0 & 0 \\ {x_{4}} & 0 & {x_{14}} & 0 & 0 & 0 &
0 & 0 \\ 0 & 0 & 0 & 0 & 0 & 0 & 0 & 0 \\ 0 & 0 & 0 & 0 & 0 & 0 & 0 & 0
\end{smallmatrix}}
 \right)$$
\end{enumerate}
Since all pairs of relevant indices lie in {\it one} orbit - the whole set $\I_3$, the
system $\es^4:\ep_k=0,\,\forall k\in \I_3$ enforces zeros on all 24 positions. This
precisely means that {\it only trivial zero solution is non-equivalent to solutions in
$\er^0,\,\er^1,\,\er^2,\,\er^3,\,\er^4$, i.e. we evaluated the whole $\er(\es_3)$ up to
equivalence.}

\bigskip
\bigskip

Our goal is to compute a set of non-equivalent normalized solutions. It will be used as
an input to a further analysis - the identification of resulting Lie algebras. We take
each solution matrix and discuss all possible combinations of zero or non-zero parameters
like in the example below.
\bigskip
\begin{example}
We take the matrix $\ep^0_2$ and let all its parameters are non-vanishing. Our question
is whether or not it is possible to normalize it to the matrix (\ref{kap}). Then the
resulting graded contractions would be isomorphic to the algebra $sl(3,\Com)$ for
arbitrary non-zero values of parameters in $\ep^0_2$. We have the normalization matrix
$\alpha$ in the form: $$\alpha=\left(
\begin{smallmatrix} 0& 0& \frac {a_{01}a_{10}}{a_{11}} & \frac
{a_{01}a_{20}}{a_{21}}  & \frac {a_{01}a_{11}}{a_{12}} &  \frac
{a_{01}a_{22}}{a_{20}} &
 \frac {a_{01}a_{12}}{a_{10}} &  \frac
{a_{01}a_{21}}{a_{22}}  \\ 0& 0&  \frac {a_{02}a_{10}}{a_{12}}  & \frac
{a_{02}a_{20}}{a_{22}}  & \frac {a_{02}a_{11}}{a_{10}} & \frac
{a_{02}a_{22}}{a_{21}} &
 \frac {a_{02}a_{12}}{a_{11}} &  \frac
{a_{02}a_{21}}{a_{20}}
\\   \frac {a_{01}a_{10}}{a_{11}} &
 \frac {a_{02}a_{10}}{a_{12}} & 0& 0&
\frac {a_{10}a_{11}}{a_{21}} &  \frac {a_{10}a_{22}}{a_{02}} & \frac
{a_{10}a_{12}}{a_{22}} &  \frac {a_{10}a_{21}}{a_{01}}
\\   \frac {a_{01}a_{20}}{a_{21}} &
 \frac {a_{02}a_{20}}{a_{22}} & 0& 0&
\frac {a_{20}a_{11}}{a_{01}} &  \frac {a_{20}a_{22}}{a_{12}} & \frac
{a_{20}a_{12}}{a_{02}} &  \frac {a_{20}a_{21}}{a_{11}}
\\   \frac {a_{01}a_{11}}{a_{12}} &
 \frac {a_{02}a_{11}}{a_{10}} &  \frac
{a_{10}a_{11}}{a_{21}} &  \frac {a_{20}a_{11}}{a_{01}} & 0& 0& \frac
{a_{11}a_{12}}{a_{20}} &  \frac {a_{11}a_{21}}{a_{02}}
\\   \frac {a_{01}a_{22}}{a_{20}} &
 \frac {a_{02}a_{22}}{a_{21}} &  \frac
{a_{10}a_{22}}{a_{02}} &  \frac {a_{20}a_{22}}{a_{12}} & 0& 0& \frac
{a_{22}a_{12}}{a_{01}} &  \frac {a_{22}a_{21}}{a_{10}}
\\   \frac {a_{01}a_{12}}{a_{10}} &
 \frac {a_{02}a_{12}}{a_{11}} &  \frac
{a_{10}a_{12}}{a_{22}} &  \frac {a_{20}a_{12}}{a_{02}} &
 \frac {a_{11}a_{12}}{a_{20}} &  \frac
{a_{22}a_{12}}{a_{01}} & 0& 0 \\   \frac {a_{01}a_{21}}{a_{22}} & \frac
{a_{02}a_{21}}{a_{20}} & \frac {a_{10}a_{21}}{a_{01}} & \frac
{a_{20}a_{21}}{a_{11}} &
 \frac {a_{11}a_{21}}{a_{02}} &  \frac
{a_{22}a_{21}}{a_{10}} & 0& 0
\end{smallmatrix}\right)$$
We only state that the system of 24 equations which was created from the matrix equality
$\ep^0_2\bullet\ \alpha=\kappa$ has a general solution in $\Com$. The matrix $\ep^0_2$
with non-zero parameters is then equivalent to the trivial solution $\kappa$ and the
corresponding graded contraction is isomorphic to $sl(3,\Com)$.
\end{example}
\bigskip
\begin{example}
\begin{enumerate}[a)]
\item  Let us take $\ep^0_2$ and put $x_{14}=0$, others parameters non-zero. The resulting matrix is denoted
as $\ep_9$ and $\nu(\ep_9)=9$ holds. The system $\ep_9\bullet\alpha=\ep_{9,I}$, where
$$\ep_{9,I}=
\begin{pmatrix} 0 & 0 & 1 & 1 & 1 & 1 & 1 & 1 \\ 0 & 0 & 1 & 1 & 1 & 1
& 1 & 1 \\ 1 & 1 & 0 & 0 & 0 & 0 & 0 & 0 \\ 1 & 1 & 0 & 0 & 0 & 1 & 0 & 1 \\ 1
& 1 & 0 & 0 & 0 & 0 & 0 & 0 \\ 1 & 1 & 0 & 1 & 0 & 0 & 0 & 1 \\ 1 & 1 & 0 & 0 &
0 & 0 & 0 & 0 \\ 1 & 1 & 0 & 1 & 0 & 1 & 0 & 0
\end{pmatrix}
$$ has a solution in $\Com$ and so $\ep_{9}$ can be normalized to $\ep_{9,I}$. To see how
the whole orbit looks like we apply all permutations corresponding to $A \in H_3$ on
$\ep_{9,I}$. The set $E_9:=\left\{(\ep_{9,I})^{\pi_A} ~\big|~A\in H_3\right\}$ has 8
elements: $$\ep_{9,I}=\begin{pmatrix} 0 & 0 & 1 & 1 & 1 & 1 & 1 & 1 \\ 0 & 0 & 1 & 1 & 1
& 1 & 1 & 1
\\ 1 & 1 & 0 & 0 & 0 & 0 & 0 & 0
\\ 1 & 1 & 0 & 0 & 0 & 1 & 0 & 1 \\ 1 & 1 & 0 & 0 & 0 & 0 & 0 & 0 \\ 1 & 1 & 0
& 1 & 0 & 0 & 0 & 1
\\ 1 & 1 & 0 & 0 & 0 & 0 & 0 & 0 \\ 1 & 1 & 0 & 1 & 0 & 1 & 0 & 0
\end{pmatrix}\q\q
\ep_{9,II}=\begin{pmatrix} 0 & 0 & 1 & 1 & 1 & 1 & 1 & 1 \\ 0 & 0 & 1 & 1 & 1 & 1 & 1 & 1
\\ 1 & 1 & 0 & 0 & 1 & 0 & 1 & 0 \\ 1 & 1 & 0 & 0 & 0 & 0 & 0 & 0
\\ 1 & 1 & 1 & 0 & 0 & 0 & 1 & 0 \\ 1 & 1 & 0 & 0 & 0 & 0 & 0 & 0 \\ 1 & 1 & 1
& 0 & 1 & 0 & 0 & 0 \\ 1 & 1 & 0 & 0 & 0 & 0 & 0 & 0
\end{pmatrix}$$ $$
\ep_{9,III}=\begin{pmatrix} 0 & 0 & 1 & 1 & 0 & 0 & 0 & 0 \\ 0 & 0 & 1 & 1 & 0 & 1 & 1 &
0
\\ 1 & 1 & 0 & 0 & 1 & 1 & 1 & 1 \\ 1 & 1 & 0 & 0 & 1 & 1 & 1 & 1
\\ 0 & 0 & 1 & 1 & 0 & 0 & 0 & 0 \\ 0 & 1 & 1 & 1 & 0 & 0 & 1 & 0 \\ 0 & 1 & 1
& 1 & 0 & 1 & 0 & 0 \\ 0 & 0 & 1 & 1 & 0 & 0 & 0 & 0
\end{pmatrix}\q\q
\ep_{9,IV}=\begin{pmatrix} 0 & 0 & 1 & 1 & 1 & 0 & 0 & 1 \\ 0 & 0 & 1 & 1 & 0 & 0 & 0 & 0
\\ 1 & 1 & 0 & 0 & 1 & 1 & 1 & 1 \\ 1 & 1 & 0 & 0 & 1 & 1 & 1 & 1 \\ 1 & 0 & 1 & 1 & 0 &
0 & 0 & 1 \\ 0 & 0 & 1 & 1 & 0 & 0 & 0 & 0 \\ 0 & 0 & 1 & 1 & 0 & 0 & 0 & 0 \\ 1 & 0 & 1
& 1 & 1 & 0 & 0 & 0
\end{pmatrix}$$ $$
\ep_{9,V}=\begin{pmatrix} 0 & 0 & 0 & 1 & 1 & 1 & 1 & 0 \\ 0 & 0 & 0 & 0 & 1 & 1 & 0 & 0
\\ 0 & 0 & 0 & 0 & 1 & 1 & 0 & 0 \\ 1 & 0 & 0 & 0 & 1 & 1 & 1 & 0
\\ 1 & 1 & 1 & 1 & 0 & 0 & 1 & 1 \\ 1 & 1 & 1 & 1 & 0 & 0 & 1 & 1 \\ 1 & 0 & 0
& 1 & 1 & 1 & 0 & 0 \\ 0 & 0 & 0 & 0 & 1 & 1 & 0 & 0
\end{pmatrix}\q\q
\ep_{9,VI}=\begin{pmatrix} 0 & 0 & 0 & 0 & 1 & 1 & 0 & 0 \\ 0 & 0 & 1 & 0 & 1 & 1 & 0 & 1
\\ 0 & 1 & 0 & 0 & 1 & 1 & 0 & 1 \\ 0 & 0 & 0 & 0 & 1 & 1 & 0 & 0
\\ 1 & 1 & 1 & 1 & 0 & 0 & 1 & 1 \\ 1 & 1 & 1 & 1 & 0 & 0 & 1 & 1 \\ 0 & 0 & 0
& 0 & 1 & 1 & 0 & 0 \\ 0 & 1 & 1 & 0 & 1 & 1 & 0 & 0
\end{pmatrix}$$ $$
\ep_{9,VII}=\begin{pmatrix} 0 & 0 & 0 & 0 & 0 & 0 & 1 & 1 \\ 0 & 0 & 0 & 1 & 1 & 0 & 1 &
1
\\ 0 & 0 & 0 & 0 & 0 & 0 & 1 & 1 \\ 0 & 1 & 0 & 0 & 1 & 0 & 1 & 1
\\ 0 & 1 & 0 & 1 & 0 & 0 & 1 & 1 \\ 0 & 0 & 0 & 0 & 0 & 0 & 1 & 1 \\ 1 & 1 & 1
& 1 & 1 & 1 & 0 & 0 \\ 1 & 1 & 1 & 1 & 1 & 1 & 0 & 0
\end{pmatrix}\q\q
\ep_{9,VIII}=\begin{pmatrix} 0 & 0 & 1 & 0 & 0 & 1 & 1 & 1 \\ 0 & 0 & 0 & 0 & 0 & 0 & 1 &
1
\\ 1 & 0 & 0 & 0 & 0 & 1 & 1 & 1 \\ 0 & 0 & 0 & 0 & 0 & 0 & 1 & 1
\\ 0 & 0 & 0 & 0 & 0 & 0 & 1 & 1 \\ 1 & 0 & 1 & 0 & 0 & 0 & 1 & 1 \\ 1 & 1 & 1
& 1 & 1 & 1 & 0 & 0 \\ 1 & 1 & 1 & 1 & 1 & 1 & 0 & 0
\end{pmatrix}$$
\item Next we take again $\ep^0_2$ and substitute $x_{18}=0$ in it and let all other
parameters be non-zero; the result is denoted as $ \ep_{9,2}$ and $\nu(\ep_{9,2})=9$. But
the matrix $\ep_{9,2}$ can be normalized to $\ep_{9,VIII}$ and thus we have discovered
that {\it the solutions $\ep_{9}$ and $\ep_{9,2}$ are equivalent.} In our sets
$\er^0,\dots$ there exists no other solution $\ep$ with the property $\nu(\ep)=9$.
Therefore we conclude that every solution in $\er(\es_3)$ with $\nu(\ep)=9$ is equivalent
to $\ep_{9,I}$ and thus only this matrix will appear on the final list of solutions as
representative of solutions with 9 zeros.
\end{enumerate}
\end{example}
\bigskip
\begin{example}
The matrix $\ep^1 \in \er^1$ contains 6 parameters $x$, these can be zero or non-zero,
and so there are $2^6=64$ cases to analyze. We take the first one and assume that all the
parameters contained in it are non-zero. We have the matrix $$\ep_{15}=\begin{pmatrix} 0
& 0 & {t_{1}} & {t_{2}} & {t_{3}} & {t_{4}} & 0 & 0 \\ 0 & 0 & {t_{7}} & 0 & 0 & {t_{10}}
& 0 & 0 \\ {t_{1}} & {t_{7}} & 0 & 0 & {t_{13}} & {t_{14}} & 0 & 0 \\ {t_{2}} & 0 & 0 & 0
& 0 & {t_{18}} & 0 & 0 \\ {t_{3}} & 0 & {t_{13}} & 0 & 0 & 0 & 0 & 0 \\ {t_{4}} &
{t_{10}} & {t_{14}} & {t_{18}} & 0 & 0 & 0 & 0
\\ 0 & 0 & 0 & 0 & 0 & 0 & 0 & 0 \\ 0 & 0 & 0 & 0 & 0 & 0 & 0 & 0
\end{pmatrix}$$
and $\nu(\ep_{15})=15$ holds. We try to normalize it to the matrix
$$\ep_{15n}=\begin{pmatrix} 0 & 0 & a & 1 & 1 & 1 & 0 & 0
\\ 0 & 0 & 1 & 0 & 0 & 1 & 0 & 0
\\ a & 1 & 0 & 0 & b & 1 & 0 & 0 \\ 1 & 0 & 0 & 0 & 0 & 1 & 0 & 0 \\ 1 & 0 & b & 0 & 0 &
0 & 0 & 0 \\ 1 & 1 & 1 & 1 & 0 & 0 & 0 & 0 \\ 0 & 0 & 0 & 0 & 0 & 0 & 0 & 0 \\ 0 & 0 & 0
& 0 & 0 & 0 & 0 & 0
\end{pmatrix}.$$
Further on, it will be clear why we have chosen $a\neq 0,\,b\neq 0$ to appear in the
matrix $\ep_{15n}$. The $24-15=9$ normalization equations are in the form:
$$\begin{tabular}{rrr}
 \parbox[l][25pt][c]{3pt}{}$\frac{a_{01}a_{10}}{a_{11}}t_1=a\q$&$\frac{a_{01}a_{20}}{a_{21}}t_2=1\q$&$\frac{a_{01}a_{11}}{a_{12}}t_3=1$\\
 \parbox[l][25pt][c]{3pt}{}$\frac{a_{01}a_{22}}{a_{20}}t_4=1\q$&$\frac{a_{02}a_{10}}{a_{12}}t_7=1\q$&$\frac{a_{02}a_{22}}{a_{21}}t_{10}=1$\\
 \parbox[l][25pt][c]{3pt}{}$\frac{a_{10}a_{11}}{a_{21}}t_{13}=b\q$&$\frac{a_{10}a_{22}}{a_{02}}t_{14}=1\q$&$\frac{a_{20}a_{22}}{a_{12}}t_{18}=1$
\end{tabular}$$
and their solution is
\begin{align*}
a_{01}= q   \\ a_{02}= q^2\frac{t_{2} t_4 }{t_{10}}
\\ a_{10}=\frac{q^2}{p}\frac{t_2 t_4}{t_{10}t_{14}}\\
a_{20}=qpt_4    \\ a_{11}=p^2\frac{t_4t_{18}}{t_3}\\          a_{22}=p ,\,\,p\neq 0
\\ a_{12}=qp^2t_4t_{18}    \\
a_{21}= q^2pt_2t_4    \\ a=\frac{ t_{1} t_{3} t_{10}}{ t_{2} t_{4} t_{7}}
\\ b=\frac{ t_{4} t_{13} t_{18}}{ t_{3} t_{10} t_{14}}
\end{align*}
where $q=\sqrt[3]{ \frac{p^3 t_{10}^2  t_{14} t_{18} }{  t_{2}^2 t_{4} t_{7} }  }$ and
$p\neq 0 $ is a parameter. We see that in general it is not possible to normalize the
matrix $\ep_{15}$ to $\ep_{15n},\,a=1,\,b=1$ because the parameters $t$ are independent.
The maximal result of normalization is a two-parameter solution $\ep_{15n}$.
\end{example}
\bigskip

We will continue in a similar vein as in the examples above - it is the most laborious
task of our problem - till we exhaust all $316$ combinations in our 20 matrices contained
in the sets $\er^0,\,\er^1,\,\er^2,\,\er^3$ and $\er^4$.

\nonumchapter{Conclusion} Using the symmetry group of the Pauli grading of $sl(3,\Com)$,
we have evaluated the set of all solutions of the corresponding contraction system up to
equivalence. For the solution of the normalization equations and for the explicit
evaluation of orbits of solutions we used the computer program Maple 8 at Centre de
recherches math\'ematiques, Montr\'eal. We proposed a sophisticated method based on
Theorem (\ref{main}) which enabled us to check all solutions in the sets
$\er^0,\,\er^1,\,\er^2,\,\er^3$ and $\er^4$ also by hands.  We remark that we did not
touch the problem of distinguishing between so called continuous and discrete graded
contractions \cite{montigny}. It is interesting to note: $$\mbox{min}
\set{\nu(\ep)\in\{1,2,3,\dots\}~\big|~\ep \in \er(\es_3)}=9$$ It turned out that there
are no solutions with less than 9 zeros (excluding the trivial solution $\kappa$).
Moreover, there are no solutions with 10, 11, 13 or 14 zeros. The complete list of
solutions is placed in the Appendix.

\nonumchapter{Appendix} The complete list of non-equivalent solutions of $\es_3$ has 180
elements. The list is divided into sections according to the number of zeros among 24
relevant parameters $\nu(\ep)$ of the solution $\ep$. In the whole list $a \neq 0,\,b\neq
0$ holds.
\bigskip
\subsection*{First trivial solution} $$\left(

 \right)$$
\smallskip
\subsection*{Solutions with 9 zeros} $$\left(
%
 \right)$$
\smallskip
\subsection*{Solutions with 12 zeros} $$\left( %
 \right)$$
\smallskip
\subsection*{Solutions with 15 zeros} $$\left(
%
 \right)$$
\smallskip
\subsection*{Solutions with 16 zeros} $$ \left(
%
 \right)$$
\smallskip
\subsection*{Solutions with 17 zeros}
$$\left(
%
 \right)$$
 \smallskip
\subsection*{Solutions with 18 zeros}
$$ \left(
%
 \right)$$
 \smallskip
\subsection*{Solutions with 19 zeros}
$$\left(
%
 \right)$$
 \smallskip
\subsection*{Solutions with 20 zeros}
$$\left(
%
 \right)$$
 \smallskip
\subsection*{Solutions with 21 zeros}
$$ \left(
%
 \right)$$
 \smallskip
\subsection*{Solutions with 22 zeros}
$$\left(
%
 \right) $$
 \smallskip
\subsection*{Solutions with 23 zeros}
$$\left( %
 \right)$$
 \smallskip
\subsection*{Second trivial solution}
$$\left( %
 \right)$$


\begin{thebibliography}{99}
\addcontentsline{toc}{chapter}{References} \setlinespacing{1.2}
\bibitem{W2}
M.~A.~Abdelmalek, X.~Leng, J.~Patera, P.~Winternitz, {\it Grading refinements in the
contraction of Lie algebras and their invariants,\/} J.~Phys.~A: Math. Gen., {\bf 29}
(1996) 7519--7543
\bibitem{W1}
    M.~Couture, J.~Patera, R.~T.~Sharp, P.~Winternitz,
{\it  Graded contractions of $sl(3,\Com)$,\/} J.~Math.~Phys. {\bf 32} (1991), 2310-2318.
\bibitem{HPP1}
{M. Havl\'\i\v cek, J. Patera, E. Pelantov{\'a}, {\em On the fine gradings of simple
classical Lie algebras}, Int. J. Mod. Phys. {\bf 12} (1997), 189--194.}
\bibitem{HPP2}
 M. Havl\'{\i}\v{c}ek, J. Patera, E. Pelantov\'a
 {\it On the maximal Abelian subgroups of diagonable
 automorphisms of simple classical Lie algebras},
 in
 ``XXI International Colloquium on Group Theoretical
 Methods in Physics'',
  World Scientific,  Singapore 1997, 116-120.
\bibitem{HPP3}
M. Havl\'{\i}\v{c}ek, J. Patera, E. Pelantov\'a {\it On Lie Gradings II}, Linear Algebra
and Its Appl. {\bf 277} (1998) 97-125.
\bibitem{HPP4}
M. Havl\'{\i}\v{c}ek, J. Patera, E. Pelantov\'a, J. Tolar, {\it Automorphisms of the fine
grading of $sl(n,\Com)$ associated with the generalized Pauli matrices}, J. Math. Phys.
Vol. 43, No.2 (2002), pp. 1083-1094
\bibitem{HPP5}
M. Havl\'{\i}\v{c}ek, J. Patera, E. Pelantov\'a, J. Tolar, {\it The fine gradings of
$sl(3,\Com)$ and their symmetries} Proceedings of XXIII International Colloquium on Group
Theoretical Methods in Physics, ed. Y. Pogosyan etal., JINR, Dubna 2001
\bibitem{HPP6}
M. Havl\'{\i}\v{c}ek, J. Patera, E. Pelantov\'a, J. Tolar, {\it Distunguished bases of
$sl(n,\Com)$ and their symmetries.} In E. Kapuscik and A. Horzela, editors, {\it Quantum
Theory and Symmetries 2,} pages 366-370, Singapore, 2002. World Scientific. Quantum
Theory and Symmetries, Proceedings of the Second International Symposium; ISBN
981-02-4887-3
\bibitem{Spanien}
F. Herranz, M. Santander, {\em The general solution of the real $\Z_2^{\otimes N}$ graded
contractions of $so(N+1)$}, J. Phys. A: Math. Gen. {\bf 29} (1996), 6643-6652.
\bibitem{jiricek}
J. Hrivn\'ak, {\em Speci\'aln\'{\i} gradovan\'e kontrakce $sl(3,\Com)$}, v\'yzkumn\'y
\'ukol FJFI \v{C}VUT, 2002.
\bibitem{montigny}
M. de Montigny and J. Patera, {\em Discrete and continuous graded contractions
  of Lie algebras and superalgebras}, J.~Phys.~A: Math. Gen. {\bf 24}
(1991), 525-549.
\bibitem{xpipa}
P. Novotn\'y, {\em Po\v{c}et prvk\r{u} a akce grupy $SL(m,\Z_n)$}, v\'yzkumn\'y \'ukol
FJFI \v{C}VUT, 2002.
\bibitem{sherp} J. Patera, R. T. Sharp and P.
Winternitz, {\it Invariants of real low dimensional Lie algebras}, J.~Math.~Phys., Vol
17, No. 6, June 1976, 986-994
\bibitem{PZ2}
J. Patera, H. Zassenhaus, {\it The Pauli matrices in $n$ dimensions and finest
 gradings of simple Lie algebras of type $A_{n-1}$,}
J. Math. Phys. {\bf 29} (1988), 665--673.
\bibitem {Pat1}
{J. Patera, H. Zassenhaus, Linear Algebra \& Its Appl. {\bf 112} (1989) 87-159}
\end{thebibliography}
\end{document}